\documentclass[twocolumn,epjc3]{svjour3}          

\RequirePackage[T1]{fontenc}
\smartqed  
\usepackage[utf8]{inputenc}
\RequirePackage{graphicx} 

\usepackage{color}   
\usepackage{hyperref}
\usepackage[switch]{lineno}
\usepackage{scalerel,amssymb}

\usepackage{ifthen}  
\newboolean{uprightparticles} 
\setboolean{uprightparticles}{false} 

\RequirePackage{graphicx}
\RequirePackage{mathptmx}      
\RequirePackage{flushend}
\RequirePackage[numbers,sort&compress]{natbib}
\usepackage[leftcaption]{sidecap}
\usepackage{graphics}
\pdfoutput=1
\usepackage{comment}
\usepackage{xcolor}
\usepackage{xurl}
\usepackage{float}
\usepackage{testhyphens}
\usepackage{hyphenat}
\usepackage{amsmath}
\usepackage{mathptmx}

\usepackage{xspace} 
\usepackage{upgreek}

\def\MagUp {\mbox{\em Mag\kern -0.05em Up}\xspace}

\ifthenelse{\boolean{uprightparticles}}%
{

 \def\PDelta      {\ensuremath{\Delta}\xspace}                 
 \def\PXi      {\ensuremath{\Xi}\xspace}                 
 \def\PLambda      {\ensuremath{\Lambda}\xspace}                 
 \def\PSigma      {\ensuremath{\Sigma}\xspace}                 
 \def\POmega      {\ensuremath{\Omega}\xspace}                 
 \def\PUpsilon      {\ensuremath{\Upsilon}\xspace}                 
 
 \def\PB      {\ensuremath{\mathrm{B}}\xspace}                 
                  
 \def\PD      {\ensuremath{\mathrm{D}}\xspace}

 \def\PK      {\ensuremath{\mathrm{K}}\xspace}

 \def\Pi      {\ensuremath{\mathrm{i}}\xspace}

}
{

 \mathchardef\PDelta="7101
 \mathchardef\PXi="7104
 \mathchardef\PLambda="7103
 \mathchardef\PSigma="7106
 \mathchardef\POmega="710A
 \mathchardef\PUpsilon="7107
                  
 \def\PB      {\ensuremath{B}\xspace}                 
                  
 \def\PD      {\ensuremath{D}\xspace}

 \def\PK      {\ensuremath{K}\xspace}

 \def\Pi      {\ensuremath{i}\xspace}

}

\makeatletter
\ifcase \@ptsize \relax
  \newcommand{\miniscule}{\@setfontsize\miniscule{4}{5}}
\or
  \newcommand{\miniscule}{\@setfontsize\miniscule{5}{6}}
\or
  \newcommand{\miniscule}{\@setfontsize\miniscule{5}{6}}
\fi
\makeatother

\DeclareRobustCommand{\optbar}[1]{\shortstack{{\miniscule (\rule[.5ex]{1.25em}{.18mm})}
  \\ [-.7ex] $#1$}}

  \def\Kbar    {{\kern 0.2em\overline{\kern -0.2em \PK}{}}\xspace}

\def\KorKbar    {\kern 0.18em\optbar{\kern -0.18em K}{}\xspace}

  \def\Dbar    {{\kern 0.2em\overline{\kern -0.2em \PD}{}}\xspace}

\def\DorDbar    {\kern 0.18em\optbar{\kern -0.18em D}{}\xspace}

\def\Bbar    {{\ensuremath{\kern 0.18em\overline{\kern -0.18em \PB}{}}}\xspace}

\def\BorBbar    {\kern 0.18em\optbar{\kern -0.18em B}{}\xspace}

  \def\Y#1S{\ensuremath{\PUpsilon{(#1S)}}\xspace}

\def\Lbar        {{\ensuremath{\kern 0.1em\overline{\kern -0.1em\PLambda}}}\xspace}
\def\LorLbar    {\kern 0.18em\optbar{\kern -0.18em \PLambda}{}\xspace}

\def\to                 {\ensuremath{\rightarrow}\xspace}

\def\AT#1     {\ensuremath{A_{\mathrm{T}}^{#1}}\xspace}           

\def\C#1      {\ensuremath{\mathcal{C}_{#1}}\xspace}                       
\def\Cp#1     {\ensuremath{\mathcal{C}_{#1}^{'}}\xspace}                    
\def\Ceff#1   {\ensuremath{\mathcal{C}_{#1}^{\mathrm{(eff)}}}\xspace}        
\def\Cpeff#1  {\ensuremath{\mathcal{C}_{#1}^{'\mathrm{(eff)}}}\xspace}       
\def\Ope#1    {\ensuremath{\mathcal{O}_{#1}}\xspace}                       
\def\Opep#1   {\ensuremath{\mathcal{O}_{#1}^{'}}\xspace}                    

\newcommand{\unit}[1]{\ensuremath{\mathrm{ \,#1}}\xspace}          

\newcommand{\tev}{\ifthenelse{\boolean{inbibliography}}{\ensuremath{~T\kern -0.05em eV}\xspace}{\ensuremath{\mathrm{\,Te\kern -0.1em V}}}\xspace}
\newcommand{\gev}{\ensuremath{\mathrm{\,Ge\kern -0.1em V}}\xspace}
\newcommand{\mev}{\ensuremath{\mathrm{\,Me\kern -0.1em V}}\xspace}
\newcommand{\kev}{\ensuremath{\mathrm{\,ke\kern -0.1em V}}\xspace}
\newcommand{\ev}{\ensuremath{\mathrm{\,e\kern -0.1em V}}\xspace}
\newcommand{\gevc}{\ensuremath{{\mathrm{\,Ge\kern -0.1em V\!/}c}}\xspace}
\newcommand{\mevc}{\ensuremath{{\mathrm{\,Me\kern -0.1em V\!/}c}}\xspace}
\newcommand{\gevcc}{\ensuremath{{\mathrm{\,Ge\kern -0.1em V\!/}c^2}}\xspace}
\newcommand{\gevgevcccc}{\ensuremath{{\mathrm{\,Ge\kern -0.1em V^2\!/}c^4}}\xspace}
\newcommand{\mevcc}{\ensuremath{{\mathrm{\,Me\kern -0.1em V\!/}c^2}}\xspace}

\def\m    {\ensuremath{\mathrm{ \,m}}\xspace}

\def\cm   {\ensuremath{\mathrm{ \,cm}}\xspace}
\def\cma  {\ensuremath{{\mathrm{ \,cm}}^2}\xspace}

\def\mma  {\ensuremath{{\mathrm{ \,mm}}^2}\xspace}

\def\ns   {\ensuremath{{\mathrm{ \,ns}}}\xspace}
\def\ps   {\ensuremath{{\mathrm{ \,ps}}}\xspace}

\def\khz  {\ensuremath{{\mathrm{ \,kHz}}}\xspace}
\def\hz   {\ensuremath{{\mathrm{ \,Hz}}}\xspace}

\def\gsim{{~\raise.15em\hbox{$>$}\kern-.85em
          \lower.35em\hbox{$\sim$}~}\xspace}
\def\lsim{{~\raise.15em\hbox{$<$}\kern-.85em
          \lower.35em\hbox{$\sim$}~}\xspace}

\def\tell1  {TELL1\xspace}
\def\ukl1   {UKL1\xspace}

\journalname{Eur. Phys. J. C}

\usepackage[normalem]{ulem}
\useunder{\uline}{\ul}{}

\makeatletter

\newcommand{\checknextarg}{\@ifnextchar\bgroup{\nolinebreak\gobblenextarg}{}}
\newcommand{\gobblenextarg}[1]{ \textsuperscript{\nolinebreak\hspace{-4pt}\mbox{\nolinebreak$^,$\nolinebreak\ref{#1}\nolinebreak}\nolinebreak} \@ifnextchar\bgroup{\gobblenextarg}{}}

\begin{document}

\title{The SHiP experiment at the proposed CERN SPS Beam Dump Facility}

\author{The SHiP Collaboration}
\institute{See on the  back for the author list \label{addr1}}




\maketitle 

\abstract{ The Search for Hidden Particles (SHiP) Collaboration has proposed a general-purpose experimental facility operating in beam-dump mode at the CERN SPS accelerator to search for light, feebly interacting particles.  In the baseline configuration, the SHiP experiment incorporates two complementary detectors. The upstream detector is designed for recoil signatures of light dark matter (LDM) scattering and for neutrino physics, in particular with tau neutrinos. It consists of a spectrometer magnet housing a layered detector system with high-density LDM/neutrino target plates, emulsion-film technology and electronic high-precision tracking. The total detector target mass amounts to about eight tonnes. The downstream detector system aims at measuring visible decays of feebly interacting particles to both fully reconstructed final states and to partially reconstructed final states with neutrinos, in a nearly background-free environment. The detector consists of a 50\m long decay volume under vacuum followed by a spectrometer and particle identification system with a rectangular acceptance of 5\,m in width and 10\,m in height. Using the high-intensity beam of 400\gev protons, the experiment aims at profiting from the $4\times 10^{19}$ protons per year that are currently unexploited at the SPS, over a period of 5--10 years. This allows probing dark photons, dark scalars and pseudo-scalars, and heavy neutral leptons with GeV-scale masses in the direct searches at sensitivities that largely exceed those of existing and projected experiments. The sensitivity to light dark matter through scattering reaches well below the dark matter relic density limits in the range from a few \mevcc up to 100\,MeV-scale masses, and it will be possible to study tau neutrino interactions with unprecedented statistics. 
This paper describes the SHiP experiment baseline setup and the detector systems, together with performance results from prototypes in test beams, as it was prepared for the 2020 Update of the European Strategy for Particle Physics. The expected detector performance from simulation is summarised at the end.}



\section{Introduction}
\label{sec:Introduction}

The SHiP Collaboration has proposed a general-purpose \linebreak intensity-frontier facility operating in beam-dump mode 
at the CERN SPS accelerator to search for  feebly interacting GeV-scale particles, here referred to as hidden sector (HS) particles, and to perform 
measurements in neutrino and in flavour physics. The SPS accelerator with its present 
performance is capable of delivering an annual yield of up to $4\times10^{19}$ protons on target 
in slow extraction of one second long spills of de-bunched beam at 400\,GeV/c, while still respecting the beam requirements of the HL-LHC and the existing SPS beam facilities. The slow extraction of de-bunched beam is motivated by the need to control the combinatorial background, and to dilute the large beam power deposited on the beam-dump target. Yields of 
${\cal O}(10^{18})$ charmed hadrons, ${\cal O}(10^{16})$ tau leptons, and ${\cal O}(10^{21})$ photons above $100~\mev$ are expected within the acceptance of the detectors in five years of nominal operation. The unprecedented yields offer a unique opportunity to
complement the world-wide program of searches for New Physics with the help of a new intensity-frontier facility that is
complementary to the high-energy and the precision frontiers.

The SHiP experiment is 
designed to both search for decay signatures of models with HS particles, such as 
heavy neutral leptons (HNL) \cite{SHiP:2018xqw}, dark photons (DP) \cite{Ahdida:2020new}, dark
scalars (DS), etc,  by full reconstruction and particle 
identification of Standard Model (SM) final states, and to search for LDM scattering signatures by the direct detection of 
recoil of atomic electrons (or nuclei) in a high-density medium \cite{SHiP:2020noy}. 
The experiment is also optimised 
to make measurements on tau neutrinos and on neutrino-induced charm production by all three species of neutrinos.

The proposal was submitted in the form of an Expression of Interest to the CERN SPS and PS committee in 2013 (CERN-SPSC-2013-024 / SPSC-EOI-010). Following the recommendation of the committee, the SHiP Collaboration proceeded with the preparation of the Technical Proposal (TP) on the detectors and on the proposed SPS experimental facility, together with an extensive report on the Physics Case, submitted in 2015 (CERN-SPSC-2015-016 / SPSC-P-350, CERN-SPSC-2015-040 / SPSC-P-350-ADD-2 and~\cite{2016SHiPPhysicsCase, Ahdida_2019}). The review of the TP concluded with a recommendation to proceed with a three-year Comprehensive Design Study, under the auspices of the CERN Physics Beyond Colliders (PBC) initiative, with the goal of submitting a proposal for the SHiP detector and the SPS Beam Dump Facility (BDF) to the 2020 European Strategy for Particle Physics Update (ESPPU) (CERN-SPSC-2019-010 / SPSC-SR-248, CERN-SPSC-2019-049 / SPSC-SR-263, CERN Yellow Reports CYRM-2020-002).

The 2020 ESPPU (CERN-ESU-014) recognised the \linebreak BDF/SHiP proposal as one of the front-runners among the new facilities investigated within CERN’s Physics Beyond Collider initiative. With regards to the cost of the baseline design of the facility, the project could not, as of 2020, be recommended for construction considering the overall recommendations of the Strategy.

In line with the ESPPU recommendations and in view of the importance of the CERN injector complex as a provider of physics programmes complementary to CERN's primary large-scale research facility, and the strong motivation behind the BDF/SHiP proposal, a continued study programme was launched in 2020. The renewed effort has initially focused on reviewing the layout of the facility and the most suitable site, including reuse of existing experimental areas at the CERN SPS, with the goal of significantly reducing the cost of the facility (CERN-SPSC-2022-009) and allow implementation to start in CERN's Long Shutdown 3.  This effort is accompanied by a re-optimisation of the experiment in order to allow integration into the alternative areas, while preserving the original physics scope and sensitivity.

The present paper describes in detail the detector and summarises the results of the detector studies and the main performance parameters as it was prepared for the 2020 \linebreak ESPPU. More details on the physics sensitivity of the experiment can be found in Ref.~\cite{SHiP:2018xqw,Ahdida:2020new,SHiP:2020noy}.



\section{Overview of the experiment}
\begin{figure*}
\centering
\includegraphics[width=0.95\textwidth]{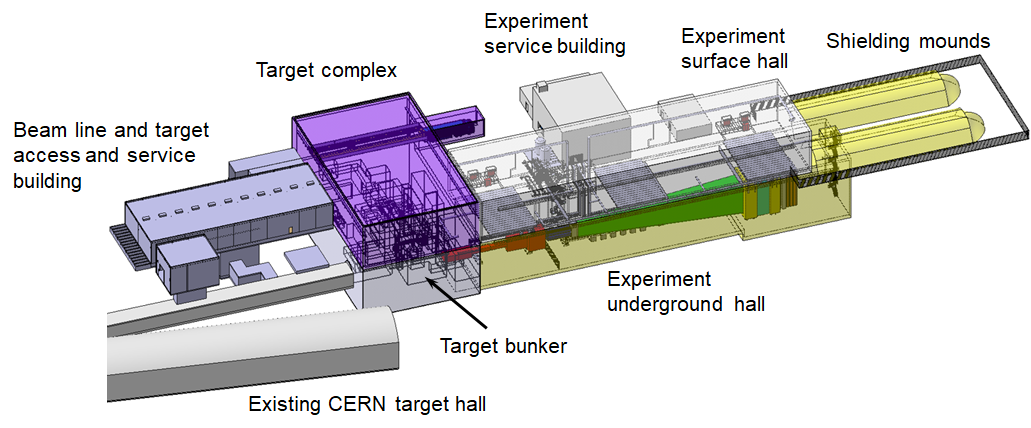}
\caption{Overview of the SPS Beam Dump Facility with the SHiP experimental area.}
\label{fig:ExperimentalAreaOverview}
\end{figure*}

\begin{figure*}
    \centering
    \includegraphics[width=0.95\linewidth]{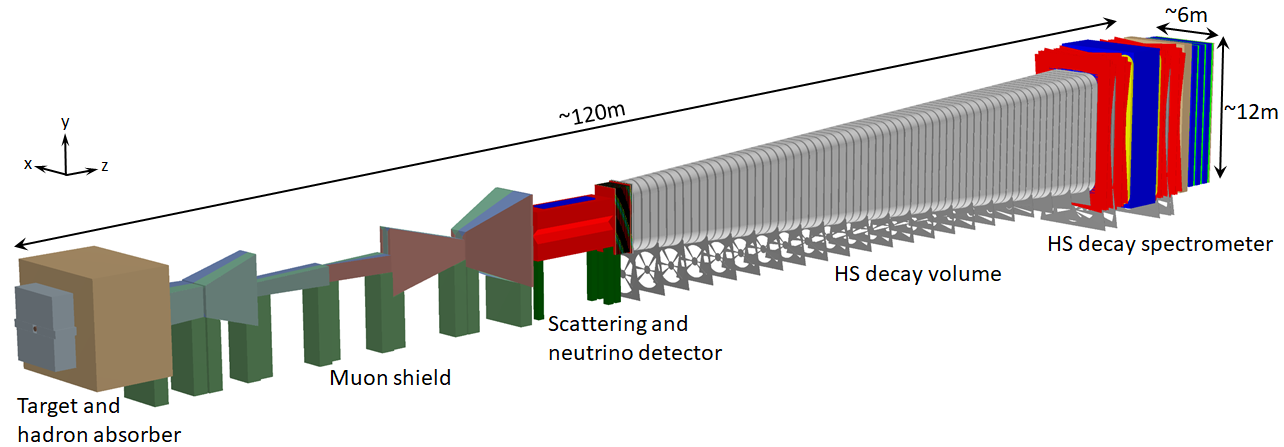}
    \caption{Overview of the SHiP experiment as implemented in the GEANT4-based~\cite{GEANT4} physics simulation.}
    \label{fig:ship_detector}
\end{figure*}

The SHiP experiment would be served by a new, short, dedicated beam line branched off the existing SPS transfer 
line to the CERN North Area. The current layout of the SPS Beam Dump Facility~\cite{Ahdida_2019} with the SHiP experimental area is shown in Fig.~\ref{fig:ExperimentalAreaOverview}, and the current layout of the SHiP experiment in Fig.~\ref{fig:ship_detector}. 

The setup consists of a high-density
proton target located in the target bunker~\cite{PhysRevAccelBeams.22.123001,PhysRevAccelBeams.22.113001,Kershaw:2018pyb}, followed by a hadron stopper and a muon shield~\cite{Akmete:2017bpl}. The target is made of blocks of a titanium-zirconium doped molybdenum alloy (TZM) in
the region of the largest deposit of energy, followed by blocks of pure tungsten. The total target depth is twelve
interaction lengths over 1.4\,m. The high atomic numbers and masses of the target material maximises the production of charm, beauty and photons as sources of HS particles, and the short interaction length ensures a high level of stopping power for pions and kaons to reduce the neutrino background. 

To control the beam-induced background from muons, the flux in the detector acceptance must be reduced from ${\cal O}(10^{11})$\hz ($>1\gev$) to less than ${\cal O}(10^{5})$\hz. Despite the aim to cover the long 
lifetimes associated with the HS particles, the detector volume should be situated as close as possible to the proton 
target due to the relatively large production angles of the HS particles.
Hence, the muon flux should be reduced over the shortest possible distance. 
To this end, an active muon shield entirely based on magnetic
deflection has been developed. The first section of the muon shield starts within the target complex
shielding assembly, one metre downstream of the target. This first section consists of a magnetic coil which magnetises the
hadron stopper with a dipole field of $\sim$1.6\,T over 4.5\,m. The rest of the muon shield
consists of six free-standing magnets, each 5\,m long, located in the upstream part of the experimental hall.

The SHiP detector consists of two complementary apparatuses, the Scattering and Neutrino Detector (SND) and the Hidden Sector Decay Spectrometer (HSDS). The SND will search for scattering of LDM and perform the neutrino measurements. It also provides normalisation of the yield for the HS particle search. The HSDS is designed to reconstruct the decay vertex of a HS particle, measure its invariant mass and provide particle identification of the decay products in an environment of extremely low background. The SND and the HSDS detectors have been designed to be sensitive to as many physics models and final states as possible, summarised in Table~\ref{tab:summary_channels}.

\begin{table*}
\centering
 \begin{tabular}{l l l} 
  & {Physics model} & {Final state}  \\ [0.5ex] 
 \hline\hline
            & HNL, SUSY neutralino                                         & $\ell^{\pm}\pi^{\mp},\,\, \ell^{\pm}K^{\mp},\,\, \ell^{\pm}\rho^{\mp}(\rho^{\mp}\rightarrow \pi^{\mp}\pi^0)$ \\
            & DP, DS, ALP (fermion coupling), SUSY sgoldstino               & $\ell^{+}\ell^{-}$ \\
 HSDS       & DP, DS, ALP (gluon coupling), SUSY sgoldstino                 & $\pi^{+}\pi^{-},\,\, K^{+}K^{-}$  \\
            & HNL, SUSY neutralino, axino                                  & $\ell^{+}\ell^{-}\nu$ \\
            & ALP (photon coupling), SUSY sgoldstino                        & $\gamma\gamma$ \\
            & SUSY sgoldstino                                               & $\pi^0\pi^0$ \\
\hline            
             & LDM                                                           & electron, proton, hadronic shower \\
SND         & $\nu_{\tau},\,\, \overline{\nu}_{\tau}$ measurements & 
            $\tau^{\pm}$ \\
            & Neutrino-induced charm production ($\nu_e, \nu_{\mu} ,\nu_{\tau}$) &  $D_s^{\pm}$, $D^{\pm}$, $D^{0}$, $\overline{D^{0}}$, $\Lambda_c^{+}$, $\overline{\Lambda_c}^{-}$ \\
 \hline
 \end{tabular}
 \caption{Summary of the physics models and final states ($\ell= e, \mu, \tau$) that the SND and the HSDS detectors are sensitive to (HNL=Heavy Neutral Lepton, DP=Dark Photon, DS=Dark Scalar, ALP=Axion-Like Particle, LDM=Light Dark Matter).}
\label{tab:summary_channels}
\end{table*}

The muon shield and the SHiP detector systems are \linebreak housed in an $\sim$120\,m long underground experimental hall at a depth of $\sim$15\,m. 


\section{Muon shield}

The free-standing magnetic muon shield~\cite{Baranov:2017chy, Akmete:2017bpl} is one of the most critical and challenging subsystems of SHiP. The baseline design relies on air-cooled warm magnets made from cold-rolled grain-oriented (CRGO) steel for an optimal compromise between a high field gradient and low input power, the latter required to reduce the space required for the coils.  The technology studies indicate that it is safe to assume an average field of 1.7\unit{T} including the magnetic-core packing factor. The optimisation of the muon shield~\cite{Baranov:2017chy,SHiP:2019gcl} has been performed with machine learning using a Bayesian optimisation algorithm and fully simulated muons from the beam-dump target by GEANT4~\cite{GEANT4}. Under the assumption that the muon shield is composed of six magnets whose geometry is described by a total of 42 parameters, the algorithm simultaneously minimises the muon background rate in the HSDS and the total mass of the shield magnet yokes.
 The input parameters describing the muon flux out of the target in the simulation were tuned to measured data with a 400\,GeV proton beam, as discussed in \cite{Ahdida:2020doo}.
The current design consists of about 600 individual packs of sheets with 50\,mm thickness, the largest with transverse dimensions of 6.6$\times$3.8\,m$^2$ and weighing about 8\,tonnes. In total, the muon shield has an overall weight of about 1400\,tonnes. The baseline for the coil consists of 9\,mm isolated stranded copper wire consolidated using an elastic compound. A table-size prototype has already been constructed and tested. 



\section{Scattering and Neutrino Detector}
\label{sec:scattering_spectrometer}

\subsection{Overview}

\begin{figure*}
\begin{center}
\includegraphics[width=0.7\linewidth]{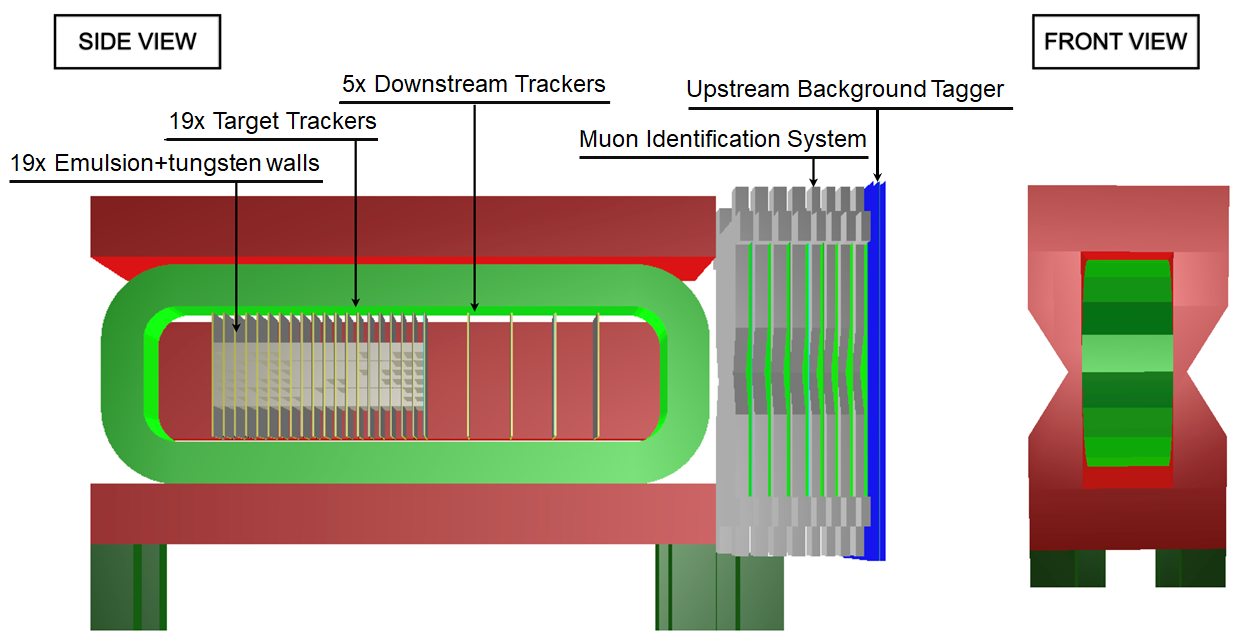}
\caption{Schematic layout of the Scattering and Neutrino Detector (SND).}
\label{fig:spectro_layout}
\end{center}
\end{figure*}

SND aims at performing measurements with neutrinos and to search for LDM. The detection of tau neutrinos and the observation for the first time of tau anti-neutrinos pose a challenge for the design of the detector, which has to fulfill conflicting requirements: a large target mass to collect enough statistics, an extremely high spatial accuracy to observe the short-lived tau lepton and a magnetic field to disentangle neutrinos from anti-neutrinos.
The optimisation of the SND layout also had to take into account constraints given by the location and the muon flux.
The overall layout, as implemented in simulation, is shown in Fig.~\ref{fig:spectro_layout}.

SND consists of a $\sim$7\,m long  detector inserted in a magnet \cite{SHiP:2019qbi} providing a 1.2\,T horizontal magnetic field, followed by a muon identification system. 
The hourglass shape of the magnet is driven by the profile of the area in the transverse plane which is cleared from the bulk of the muon flux, which in turn defines the allowed region for the detector. The neutrino flux and neutrino energy decrease with larger polar angles and this produces a radial dependence of the neutrino interaction yield, 
favouring the development of a narrow and long emulsion target.
 


The magnet hosts the emulsion target, interleaved with target tracker planes, and a downstream tracker.
The emulsion target has a modular structure: the unit cell consists of an emulsion cloud chamber (ECC) made of tungsten plates interleaved with nuclear emulsion films, followed by a compact emulsion spectrometer (CES) for the momentum and charge sign measurement of particles produced in neutrino interactions. The ECC bricks are arranged in walls alternated with target tracker planes, providing the time stamp of the interactions occurring in the target. The downstream tracker is made of three target tracker planes separated by $\sim$50\,cm air gaps. It is used to measure the charge and momentum of muons exiting the target region, thus extending significantly the detectable momentum range of the CES. The downstream tracker planes also help to connect the tracks in the emulsion films with the downstream SND muon identification system. 

The SND muon identification system is made of a sequence of iron filters and resistive plate chamber (RPC) planes, totalling about two metres in length. As neutrino interactions in the iron can generate background for the HS decay search through the production of long-lived neutral particles entering the downstream HSDS decay volume and mimicking signal events, the downstream part of the muon identification system also acts as an HSDS upstream background tagger (UBT).

\subsection{SND emulsion target}

\begin{figure}
\centering
\includegraphics[width=1.0\linewidth]{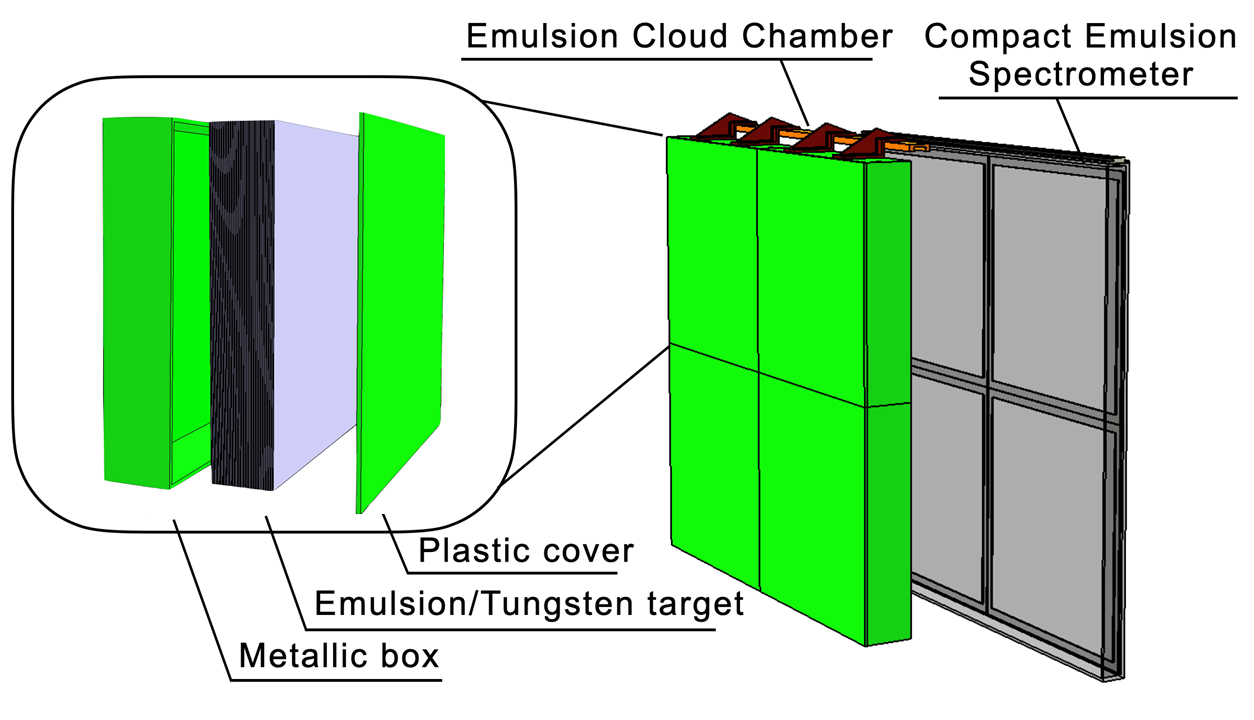}
\caption{Layout of the emulsion target and closeup view of one emulsion brick wall of four cells, each containing an ECC and a CES.}
\label{fig:target_layout}
\end{figure}

The emulsion target is in the current baseline made of 19 emulsion brick walls and 19 target tracker planes. The walls are divided in 2$\times$2 cells, each with a transverse size of 40$\times$40 cm$^2$, containing ECC and a CES as illustrated in Fig.~\ref{fig:target_layout}.

\begin{figure}
\centering
\includegraphics[width=1.0\linewidth]{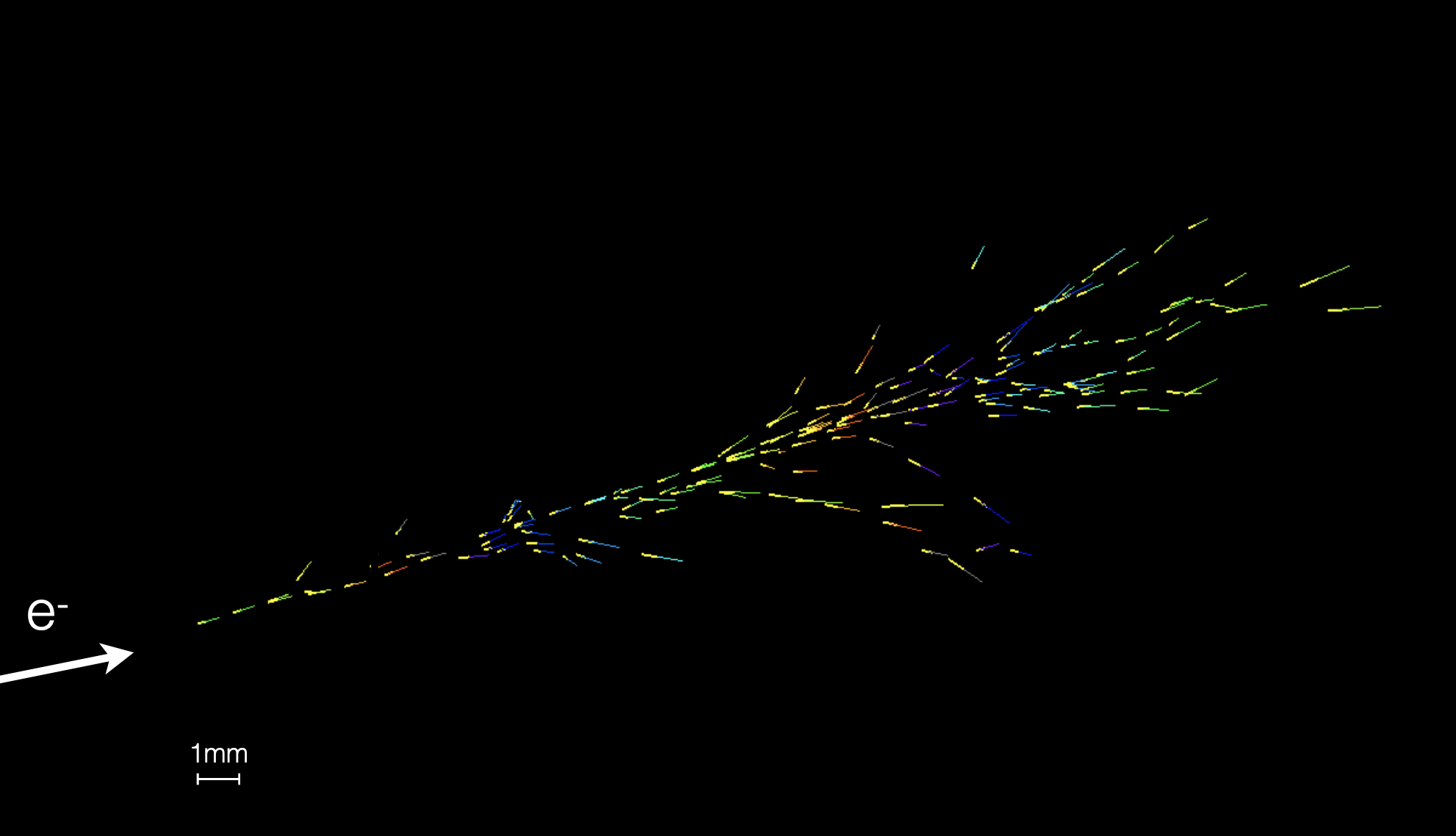}
\caption{Simulated electromagnetic shower induced in an ECC by a 6\,\gevc electron. The picture shows the $e^+e^-$-track segments reconstructed in the emulsion films.}
\label{fig:ecc_shower_display}
\end{figure}

The ECC technology makes use of nuclear emulsion \linebreak films interleaved with passive absorber layers to build up a tracking device 
with sub-micrometric position and milliradian angular resolution, as demonstrated by the OPERA experiment~\cite{Acquafredda:2009zz}. 
The technique allows detecting tau leptons~\cite{Agafonova:2018auq} and charmed hadrons~\cite{Agafonova:2014khd} by disentangling their production and decay vertices. It is also suited for LDM detection through the direct observation of the scattering off electrons in the absorber layers. The high spatial resolution of the nuclear emulsion allows measuring the momentum of charged particles through the detection of multiple Coulomb
scattering in the passive material~\cite{OPERA:2011aa}, and identifying electrons by observing electromagnetic showers in the brick~\cite{Agafonova:2018dkb}.
Nuclear emulsion films are produced by Nagoya University in collaboration with the Fuji Film Company and by the Slavich Company in Russia. 

An ECC brick is made of 36 emulsion films with a transverse size of  40$\times$40\cma, interleaved with 1\,mm thick tungsten layers. Tungsten has been chosen in place of lead as in the OPERA experiment  for its higher density and for its shorter radiation length and smaller Moli\`ere radius in order to improve the electromagnetic-shower containment. The resulting brick has a total thickness of $\sim$5 cm, corresponding to $\sim$10\,$X_0$, and a total weight of $\sim$100\,kg. The overall target weight with 19 walls of 2$\times$2 bricks is about 8\,tonnes. With the estimated background flux, the emulsion films must be replaced twice a year in order to keep the integrated amount of tracks to a level that does not spoil the reconstruction performance. The films are analysed by fully automated optical microscopes \cite{Arrabito:2006rv,Armenise:2005yh}. The scanning speed, measured in terms of film surface per unit time, was significantly increased in recent years \cite{Alexandrov:2015kzs,Alexandrov:2016tyi,Alexandrov:2017qpw,Alexandrov:2019dvd}.  The current scanning speed and the availability of several tens of microscopes within the collaboration makes it possible to scan the whole emulsion film surface over a time scale of six months.

\begin{figure}
\centering
\includegraphics[width=1.0\linewidth]{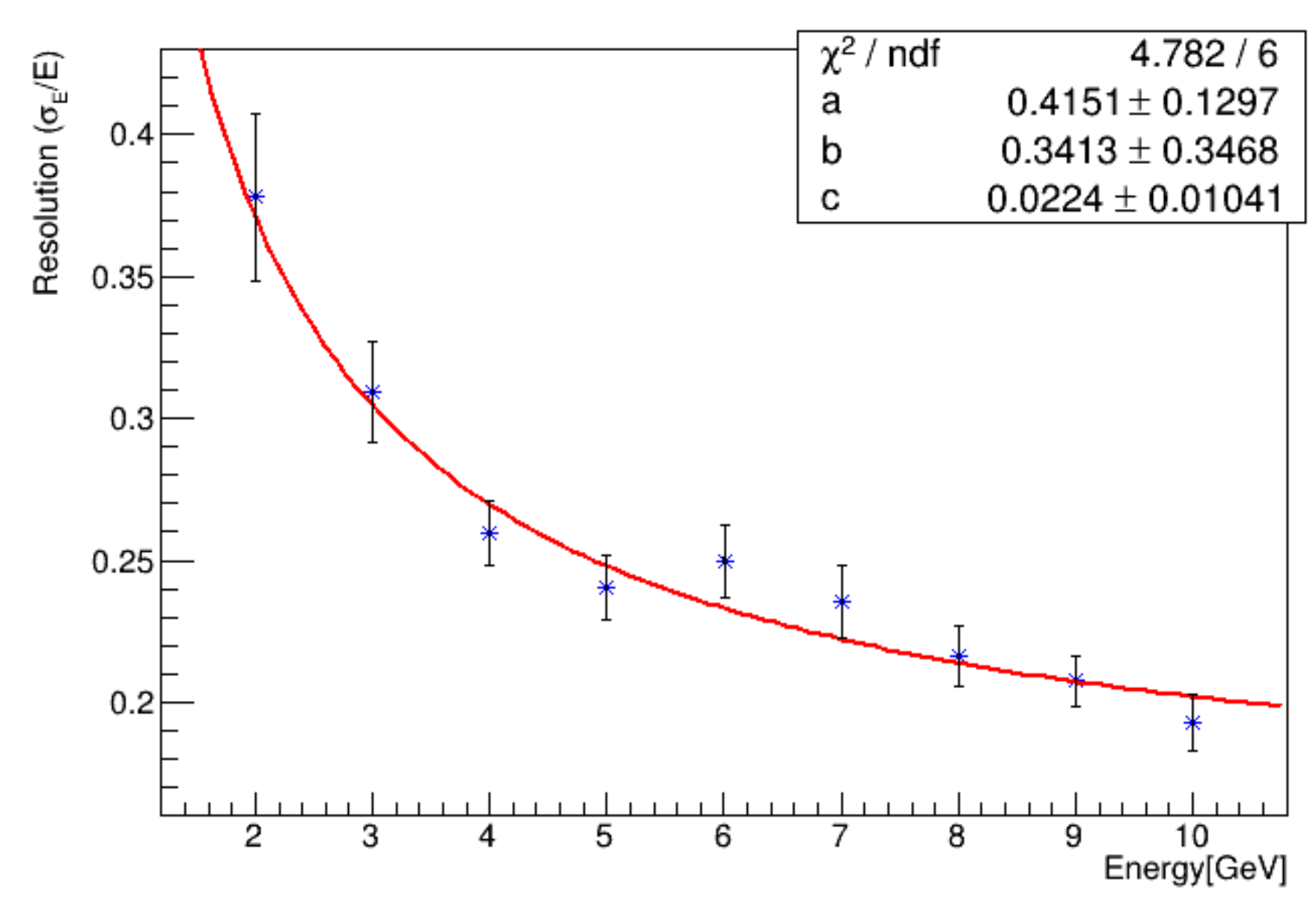}
\caption{Energy resolution for electrons using the ECC as sampling calorimeter, as estimated with MC simulations. The function $ \sigma_E/E = \sqrt{(a/\sqrt{E})^2 + (b/E)^2 + c^2}$
was used for the fit.}
\label{fig:ecc_shower_energy}
\end{figure}

The performance of the ECC in terms of  electromagnetic shower identification and energy measurement was studied with Monte Carlo (MC) simulations for different electron energies. The ECC can be used as a sampling calorimeter, with the number of track segments being proportional to the energy of the electron (see Fig.~\ref{fig:ecc_shower_display}). A new approach based on machine learning techniques was developed and an energy better than 25\% was achieved for energies higher than 4\gev, as reported in Fig.~\ref{fig:ecc_shower_energy}.

\begin{figure}
\centering
\includegraphics[width=0.8\linewidth]{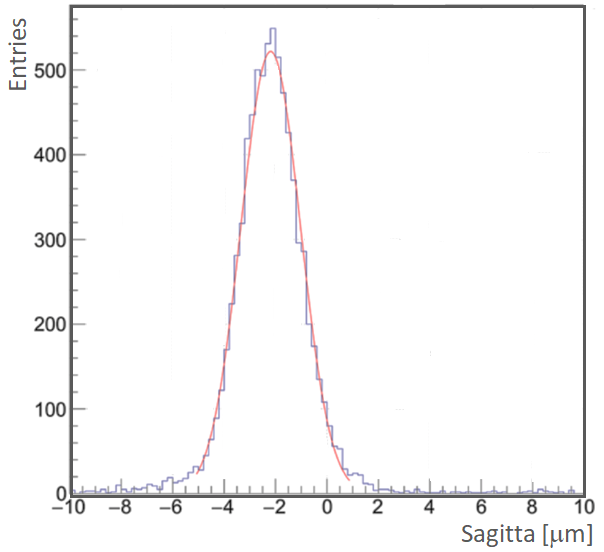}
\caption{Measured sagitta for 10\,\gevc pions in a CES prototype tested in 2017 at the CERN PS.}
\label{fig:target_sagitta}
\end{figure}

The CES modules aim at measuring the electric charge of hadrons produced in tau lepton decays, thus providing the unique feature of disentangling $\nu_\tau$ and $\overline{\nu}_\tau$ charged-current interactions also in their hadronic decay channels. It complements the use of the ECC in the momentum measurement for hadrons and soft muons which are emitted at large angles and which do not reach the downstream tracker. The basic structure of the CES is made of three emulsion films interleaved by two layers of low density material. The emulsion films belonging to the CES will need more frequent replacements than those of the ECCs since reconstruction requires a lower level of background tracks. The replacement frequency is part of the current investigations. 
The CES concept was demonstrated in 2008 \cite{Fukushima:2008zzb}.
A new version with air gaps was designed and tested in 2017 at the CERN PS. The air gap was made by a 15\,mm-thick poly(methyl methacrylate) (PMMA) hollow spacer placed between consecutive emulsion films. 
Different emulsion-film prototypes were tested in order to identify the support for the emulsion which minimises local deformations. Results show that the use of a 500\,$\upmu$m-thick glass base induces deformations on the emulsion surface which are five times smaller than the 175\,$\upmu$m-thick PMMA base typically used.
Results obtained with the CES made with a glass base are very promising. The distributions of the measured sagitta along the $x$-axis for 1 and 10\,\gevc pions show gaussian peaks with $\sigma$ of 10.2 and 1.15\,$\upmu$m, respectively (Fig.~\ref{fig:target_sagitta}).
A momentum resolution of $\sim$30\% up to 10\,\gevc momenta was achieved for the first time. 

\subsection{SND target tracker and downstream tracker}

A system of electronic detectors in combination with the ECCs is needed in order to time stamp the events reconstructed inside the bricks and to connect the emulsion tracks to those reconstructed in the downstream tracker and in the SND muon identification system.

In the baseline configuration, the SND employs 19 target tracker planes within the emulsion target, with the first one acting as a veto for charged particles entering the emulsion target. The three additional downstream tracker planes located immediately after the emulsion target measure momentum and charge for long tracks, in particular muons. The distance between two downstream tracker planes is 50\,cm without any material interposed between them. 

The baseline technology for the SND tracker systems consists of a scintillating-fibre tracker (SciFi). Its main characteristics are: high granularity tracking with a spatial resolution of 50\,$\upmu$m over a surface of $\sim$1\,m$^2$, single plane time resolution of 400\,ps, high detection efficiency of $>$99.5\%, and insensitivity to magnetic field. The active detector planes are composed of six fibre layers glued together to form fibre mats of $\sim$1.5\,mm thickness. The fibre mats are glued on supports made of carbon fibre honeycomb structures, forming large detector surfaces with dead-zones of less than \linebreak 500\,$\upmu$m along the borders with adjacent fibre mats. The total thickness of an $x-y$ plane is less than 15\,mm. The spatial resolution has been measured to be 36\,\textmu m by LHCb.   


The active area of each plane is $917$\,$\times$\,$1440$\,mm$^{2}$. The dimensions exceed those of the ECC and CES in order to track particles emitted at large angles in several consecutive walls downstream of the ECC in which the interaction occurred. Additionally, a signal-cluster shape analysis allows a modest single-hit angular resolution which helps to resolve possible combinatorial ambiguities. 


\subsection{SND muon identification system}
\label{sec:rpcs}
  
The  SND muon identification system is designed to identify with high efficiency the muons produced in neutrino interactions and $\tau$ decays occurring in the emulsion target. 
The system consists of eight hadron filters of iron interleaved with tracking planes instrumented with RPCs, plus three additional downstream layers based on multigap resistive plate chamber (MRPC) planes. The MRPC planes also act as the upstream background tagger for the HS decay searches, and is described in Section~\ref{sec:ubt}.
The four most upstream iron layers are 15\,cm thick. The downstream layers are 10\,cm thick in order to have a better tracking performance. 

 Each RPC plane is made of three gaps with an active area of 1900$\times$1200\,mm$^2$ each. The RPC planes are read out by means of orthogonal-strip panels with $\sim$1\,cm pitch.  The overall transverse dimension of one plane is 4290 $\times$ 2844\,mm$^2$. The planes are staggered by $\pm$10\,cm to compensate for the acceptance loss due to the dead areas between adjacent gaps in the same plane. Due to the significant rate of beam-induced particles impinging on the muon system, the RPCs will be operated in avalanche mode.



The front-end boards house two ASICs with eight input channels each, performing amplification and  discrimination of signals, and an on-board FPGA for data time-stamping, zero suppression and serialization. Each RPC plane are read out by 38 boards. 
The current option for the ASICs is the FEERIC chip~\cite{Dupieux:2014sca}, developed by the ALICE Collaboration.

\begin{figure}
    \centering
    \includegraphics[width=0.99\linewidth]{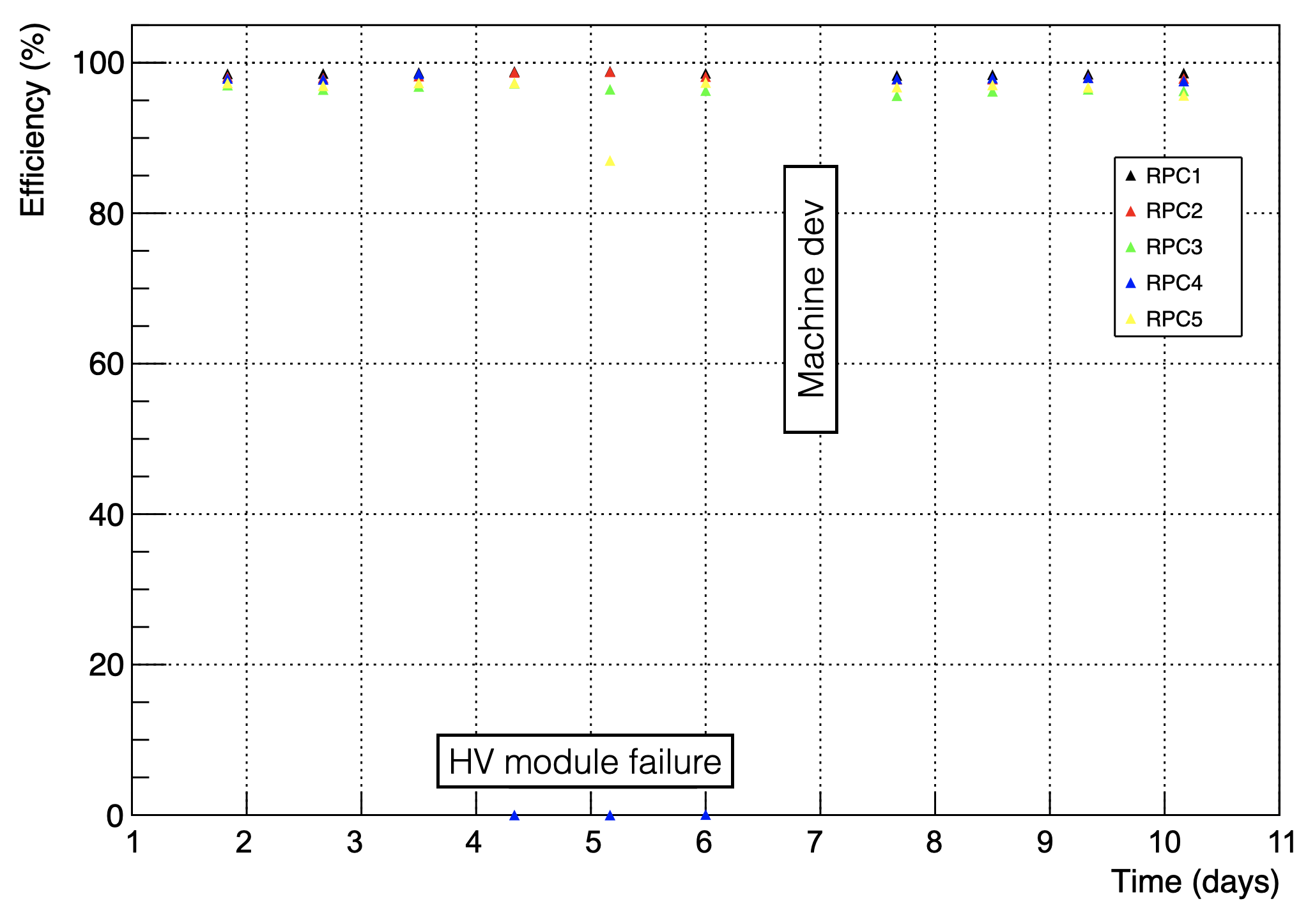}
    \caption{ Efficiency of the five RPC stations used  in the muon-flux and charm-production measurement. During the run, RPC chamber 4 suffered a problem with the high-voltage module during three days and SPS had a day of machine development studies.}
    \label{fig:SND_eff}
\end{figure}

Five prototype RPC detectors, each consisting of a 2\,mm-wide gas gap with 2\,mm-thick Bakelite electrodes and an active area of 1900$\times$1200\,mm$^2$, were constructed and used in CERN H4 beam line for the SHiP muon-spectrum and charm-production measurements in July 2018~\cite{Ahdida:2020doo}. The chambers, read out by two panels of orthogonal copper strips with a pitch of 10.625\,mm were operated in avalanche mode with a standard gas mixture of C$_2$H$_2$F$_4$/C$_4$H$_{10}$/SF$_6$ in volume ratios of 95.2\%/4.5\%/0.3\%, respectively. As shown in Fig.~\ref{fig:SND_eff}, the chambers showed efficiencies above 98\%. 

\section{Hidden Sector Decay Spectrometer}

\subsection{Overview}

HS particles are typically endowed with long lifetimes compared to most unstable SM particles, making the beam-dump configuration with a large fiducial volume particularly suitable. The decay volume of SHiP is located immediately downstream of the SND at $\sim$45\,m from the centre of the proton target. The geometry and dimensions of the detector volume that define the decay acceptance have been obtained by an optimisation based on a wide range of physics models and particle masses, the performance of the muon shield, and the $5\m\times 10\m$ aperture of SHiP's spectrometer. The resulting decay volume delineates a pyramidal frustum with a length of 50\m and upstream dimensions of $1.5\m\times4.3\m$. 

The signature of a HS particle decay consists of an isolated decay vertex within the decay volume and no associated activity in the tagger system surrounding the decay volume. For fully reconstructed signal decays, where all particles coming from the decaying hidden particle are reconstructed in the spectrometer,  the total momentum vector of the decay vertex extrapolates back to the proton interaction region. In partially reconstructed final states with one or more missing particles, e.g. as in one of the decay modes of heavy neutral leptons, HNL$\to \mu^+\mu^- \nu$, the total momentum vector of the decay vertex extrapolates back to a larger region around the proton interaction region. Particle identification of the decay products is used in order to further suppress background and to allow characterising a signal and associating it with physics models. 

The search for visible decays requires an extremely low and well-controlled background environment with a highly redundant detector system.
The dominant sources of background are neutrino-induced and muon-induced inelastic interactions in the detector or surrounding materials, and random combinations of reconstructed residual muons.
Backgrounds originating from cosmic muons have been demonstrated to be negligible (CERN-SPSC-2015-016 / SPSC-P-350). 

In order to suppress the background from neutrinos interacting with air in the fiducial volume, the decay volume is maintained at a pressure of $<10^{-3}$\,bar by means of a vacuum vessel. In this configuration, neutrino interactions mainly occur in the vessel walls.
Neutral long-lived background particles from these neutrino interactions that decay in the decay volume can be rejected by the reconstructed impact parameter at the beam-dump target. The impact parameter at the proton target is very powerful in rejecting all background sources to fully reconstructed final states. Partially reconstructed final states are more challenging to discriminate from the background since they require a looser criterion on the impact parameter. To further ensure that signal candidates are not produced by neutrino or muon interactions in the upstream SND or in the decay volume walls, the decay volume is completely covered by a high-efficiency background tagger system which is capable of detecting the charged particles produced in the interactions of muons and neutrinos. Requiring no local activity in the background tagger in time with the decay candidate is very efficient in suppressing residual background events in the partially reconstructed modes. Timing coincidence with high resolution is used to reject background from fake decay vertices formed by random combinations of muon tracks. The particle identification system provides further background rejection.


\begin{figure*}
    \centering
    \includegraphics[width=0.75\linewidth]{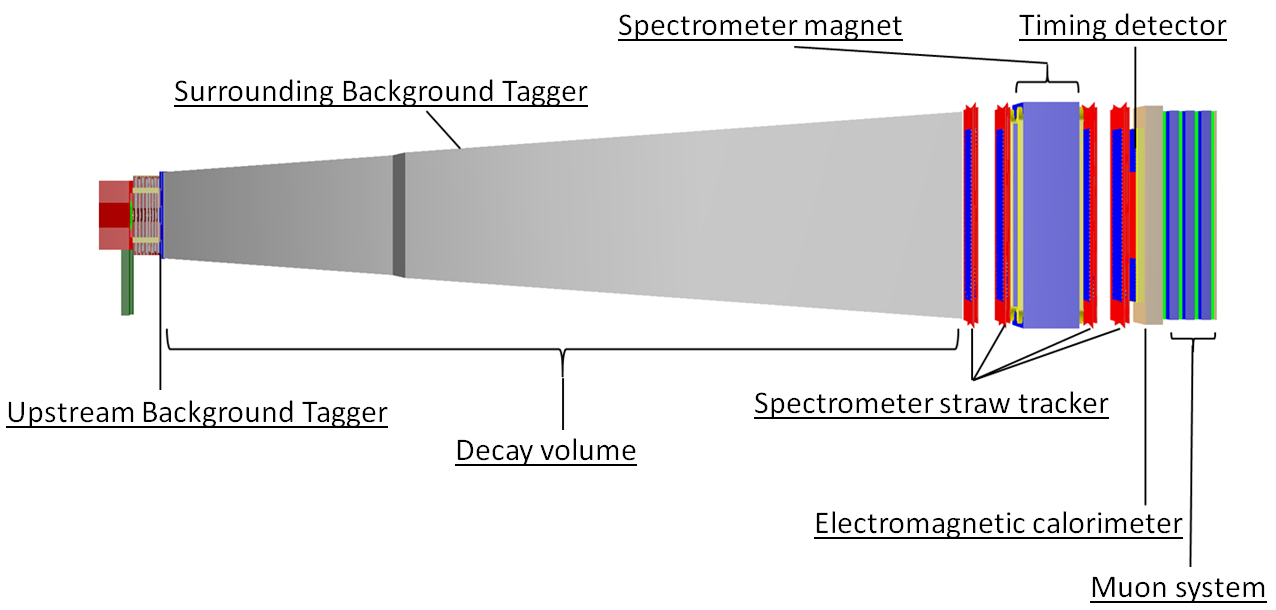}
    \caption{Schematic layout of the HS Decay Spectrometer (HSDS).}
    \label{fig:HS_DecayVolume}
\end{figure*}

The various sub-systems of the HSDS are indicated in Fig.~\ref{fig:HS_DecayVolume} and are described in details in the following.




\subsection{HSDS vacuum vessel and magnet}

\begin{figure*}
    \centering
    \includegraphics[width=0.7\linewidth]{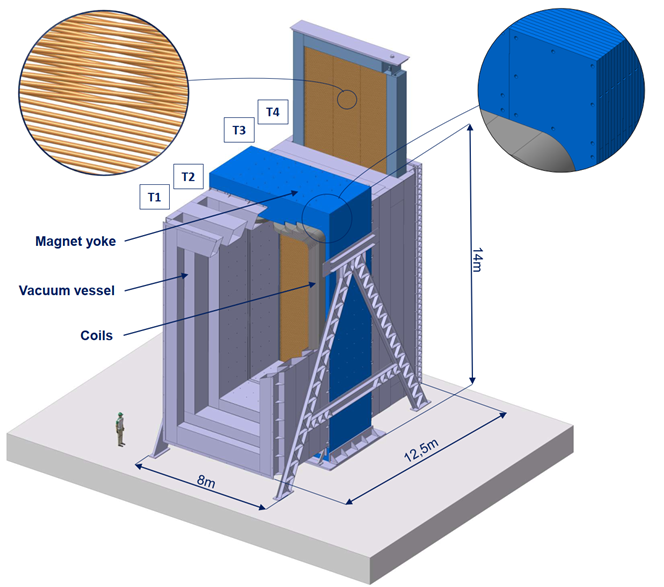}
    \caption{Layout of the spectrometer section, with the four top openings
in the vacuum vessel for the insertion of the straw tracker stations (T1 -- T4),
and the surrounding spectrometer magnet. The top-left magnified view
illustrates the orientation of the tracker straw tubes, and the top right
zooms on the magnet yoke plates with through-holes for assembling with rods.}
    \label{fig:HS_SpectrometerSection}
\end{figure*}

In order to avoid material between the decay volume and the spectrometer straw tracker (Section~\ref{sec:SST}), the HSDS vacuum vessel logically consists of two connected parts, the volume in which a decay vertex is accepted and the spectrometer section (Fig.~\ref{fig:HS_SpectrometerSection}). The spectrometer section runs through the spectrometer magnet and includes four tracker stations, two stations upstream and two downstream of the magnet. An upstream and a downstream end-cap close off the ends of the vacuum vessel. The two sections and the end-caps are connected by bolted flanges. To ensure that signal candidates are not produced by neutrino or muon interactions in the upstream SND or the walls of the vacuum vessel, the decay volume is completely covered by the high-efficiency upstream background tagger (UBT) (Section~\ref{sec:ubt}) and a surrounding background tagger (SBT) (Section~\ref{sec:sbt}) which are capable of detecting the charged particles produced in the interactions.

The decay-volume wall structure has been optimised~\cite{DecayVolumeOptimisation} in order to be as thin and light as possible, and to incorporate the SBT liquid-scintillator detector in compartments with dimensions of $0.80 \times 1.20 - 1.5$\,m$^2$. The final design consists of a double-wall structure with an internal skeleton of azimuthal beams and longitudinal strengthening members entirely based on S355JO(J2/K2)W Corten steel.

The spectrometer section of the vacuum volume is constructed from austenitic stainless steel and is mechanically supported by the magnet yoke. The tracker is inserted into the vacuum by a top-loader system including a flange and cover.  This results in a total vacuum-vessel volume of \linebreak $\sim$\,2040\,m$^3$. 

The vacuum volume downstream end-cap is located just behind the last tracker station upstream of the timing and the particle identification detectors. The baseline design of both the upstream and downstream end-caps is based on a flat panel weld from a vertical stack of extruded profiles of aluminium alloy of type 6060. The material budget is equivalent to $0.8$ radiation lengths in order to minimise the risk of neutrino and muon interactions in the upstream end-cap and not spoil the calorimeter performance after the downstream end-cap.

The HSDS magnet is based on a warm \linebreak conventional magnet, see Fig.~\ref{fig:HS_SpectrometerSection}. It is required to have a physics aperture of $5\times 10$\,m$^2$ and provide a vertical bending power of about $\sim$0.65\,Tm over the distance between the upstream and the downstream tracking stations. As the magnet aperture is limited in the horizontal plane by the region cleared from the beam-induced muon flux, the choice of the horizontal field orientation is motivated by the shorter field gap. The coils are made from a square-shape hollow aluminum conductor with transverse dimensions of $50\times 50\,\mma$ and a bore hole of 25\,mm for water cooling.
The yoke is built from packs of 50\,mm thick sheets of AISI\,1010 steel. The pack is assembled in a brick-laying fashion around the corners. In terms of aperture, 100\,mm has been reserved all around the physics aperture to accommodate the vacuum vessel and its anchoring within the yoke. The result is a 1155-tonne yoke with two vertical coil packs of 25\,tonnes each. Simulations show that the required magnetic performance can be obtained with a current density of 1.5\,A/\mma and an excitation current of 3000\,A, resulting in a total power consumption of $\sim$\,1.1\,MW.

\subsection {HSDS background taggers}


\begin{figure}
    \centering
    \includegraphics[width=0.7\linewidth]{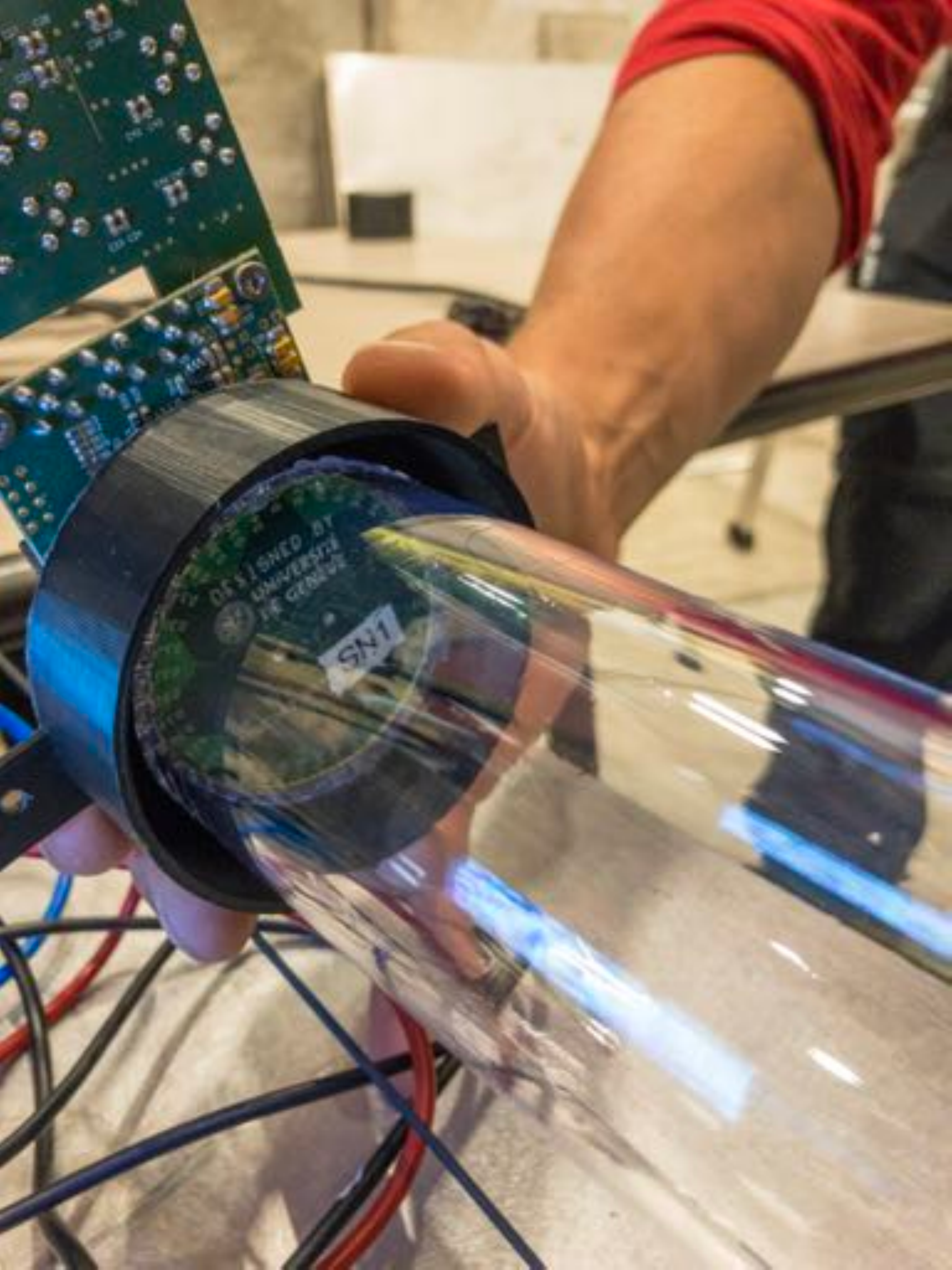}
    \caption{Coupling of the SBT wavelength-shifting optical module (WOM) tube to a 40-SiPM ring-array printed circuit board.}
    \label{fig:HS_SBT_WOMwithinSiPMarray}
\end{figure}

\begin{figure}
    \centering
    \includegraphics[width=0.95\linewidth]{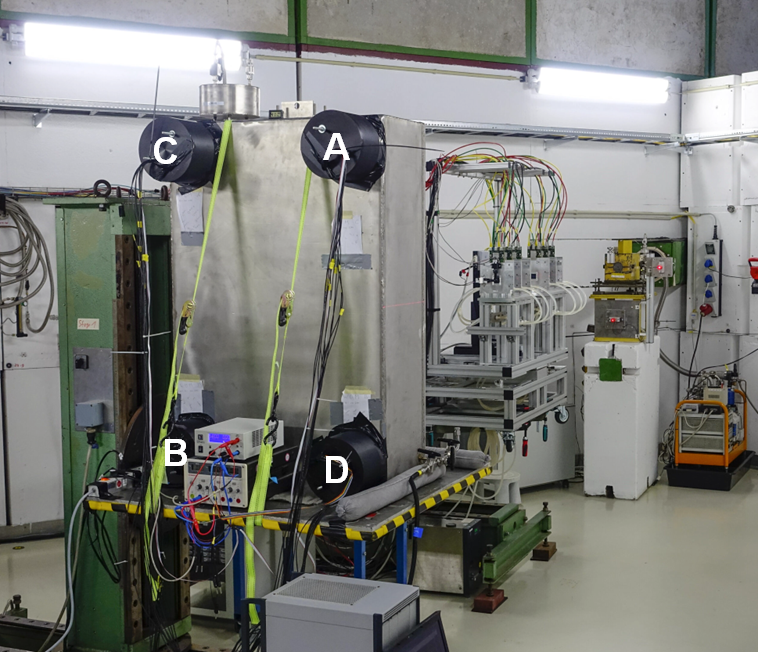}
    \caption{Experimental setup of the SBT liquid-scintillator cell prototype, equipped with four WOMs (at positions labelled A, B, C, D) at a DESY electron test beam in 2019.}
    \label{fig:HS_SBT_TestbeamSetup2019}
\end{figure}

\begin{figure}
    \centering
    \includegraphics[width=0.99\linewidth]{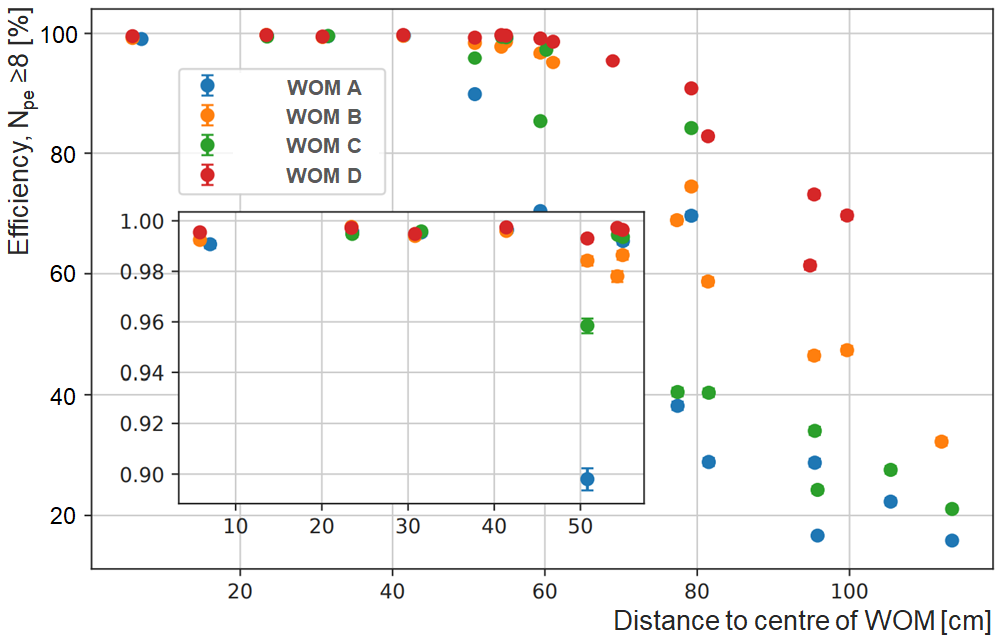}
    \caption{Detection efficiency by single SBT WOM for muons as a function of the distance between the muon beam and the WOM. A, B, C, D refer to the four WOMs positioned as indicated in Fig.~\ref{fig:HS_SBT_TestbeamSetup2019}. The insert shows the relevant region magnified.}
    \label{fig:HS_SBT_Efficiency_vs_position_8_allWOMs_muons}
\end{figure}

\subsubsection{UBT}
\label{sec:ubt}

The UBT covers the front-cap window of the vacuum vessel and is designed to tag the time and position of charged particles produced by neutrino interactions in the passive material of the SND muon identification system. A good time resolution, around 300\,ps, is needed. Having a position resolution of the order of a few millimeters, this system also provides position information for muon tracks, complementing the RPCs in the SND muon identification. The envisaged technology is similar to the technology used in the HSDS timing detector based on MRPCs described in Section~\ref{subsec:TDMRPC}, but given the less stringent requirements on the time resolution, a multi-gap RPC structure with only two gas gaps of $1$ mm width is sufficient. To cover the HSDS vacuum vessel entrance window, each tagger plane is made of five MRPC modules (of about 2070$\times$1020\,mm$^{2}$) arranged vertically with an overlap of $10$\,cm in order to create layers without dead regions.

This technology has been successfully tested~\cite{R3B_p,R3B_n} and is being used in the framework of other projects~\cite{tragaldabas_2014}. Experimental tests performed with a single MRPC with an active area of $1.5$×$1.2$~m$^2$ have shown 90\% efficiency on the whole surface (limited by the pick-up electrode, which covers 90\% of the detector active area), and about 300\,ps time resolution.

\subsubsection{SBT}
\label{sec:sbt}

The SBT must be capable of detecting charged particles either entering the vacuum vessel side walls from outside, or produced in inelastic interactions of muons and neutrinos in the vessel walls. The baseline option to cover the top-, the bottom-, and the side-walls of the vacuum vessel is using a state-of-the-art liquid scintillator (LS-SBT) consisting of linear alkylbenzene (LAB) together with 2.0\,g/l diphenyl-oxazole (PPO) as the fluorescent. This technology provides a high detection efficiency and good time resolution at a reasonable cost.
    
The LS-SBT is sub-divided into individual cells integrated into the support structure of the vacuum vessel. This results in cell sizes of 80\,cm in the longitudinal direction, and between $\sim$80\,cm and $\sim$150\,cm in the azimuthal direction, depending on the location along the vacuum vessel. The thickness of the liquid-scintillator layer volume is about 30\,cm, varying slightly along the length of the vacuum vessel, making up a total volume of $\sim$300\,m$^3$.

Each cell of the SBT is read out by two wavelength-shifting optical modules (WOM 
) made from PMMA tubes (length: 23\,cm, diameter: 6\,cm, wall thickness: 3\,mm) that are dip-coated with a wavelength-\-shifting dye (77.31\% \linebreak toluene, 22.29\% paraloid B723, 0.13\% bis-MSB and 0.27\% \linebreak p-terphenyl~\cite{Hebecker:2016mrq}) 
The WOMs absorb scintillation light in the range of 340\,nm -- 400\,nm. 
The re-emitted photons with wavelengths above 400\,nm are guided by the WOM tube to a ring of 40 SiPMs, each with a $3\times 3$\,mm$^2$ area, directly coupled to the WOM tube (see Fig. \ref{fig:HS_SBT_WOMwithinSiPMarray}). There are O(4000) WOMs for the whole SBT. Beam test measurements in 2017 demonstrated the principle with a small-scale prototype of WOM-equipped liquid-scintillator cell~\cite{Ehlert:2018pke}. Combining two WOM signals, a time resolution of 1\,ns and a homogeneity of the detector response over the detector
volume within 20\% are achieved. Further beam tests in 2018 and 2019 with
a prototype of $120\times 80 \times 30$\,cm$^3$ (see Fig.~\ref{fig:HS_SBT_TestbeamSetup2019}) achieved a detection efficiency for charged particles depositing at least 45\,MeV, corresponding to a minimum-ionising particle passing about 30 cm of liquid scintillator, close to 99.9\% for distances between the passing particle and the WOM up to about 60\,cm (see Fig.~\ref{fig:HS_SBT_Efficiency_vs_position_8_allWOMs_muons}). Tests with a small-scale cell using a liquid scintillator purified by Al$_{2}$O$_{3}$ columns as e.g. described in \cite{Benziger:2007aq} were performed and show an increase of 20\% in light yield compared to measurements with an unpurified liquid scintillator. GEANT4-based photon-\-transport simulations show that covering the cell walls with acrylic BaSO$_4$-coating is expected to increase the detected light yield by a factor of about five. As a result, the detection technique is well-suited to achieve 99.9\% detection efficiency over large-area liquid-scintillator filled cells for energy depositions even below 45\,MeV. 

\subsection{HSDS straw tracker}
\label{sec:SST}

\begin{figure}
    \centering
    \includegraphics[width=0.95\linewidth]{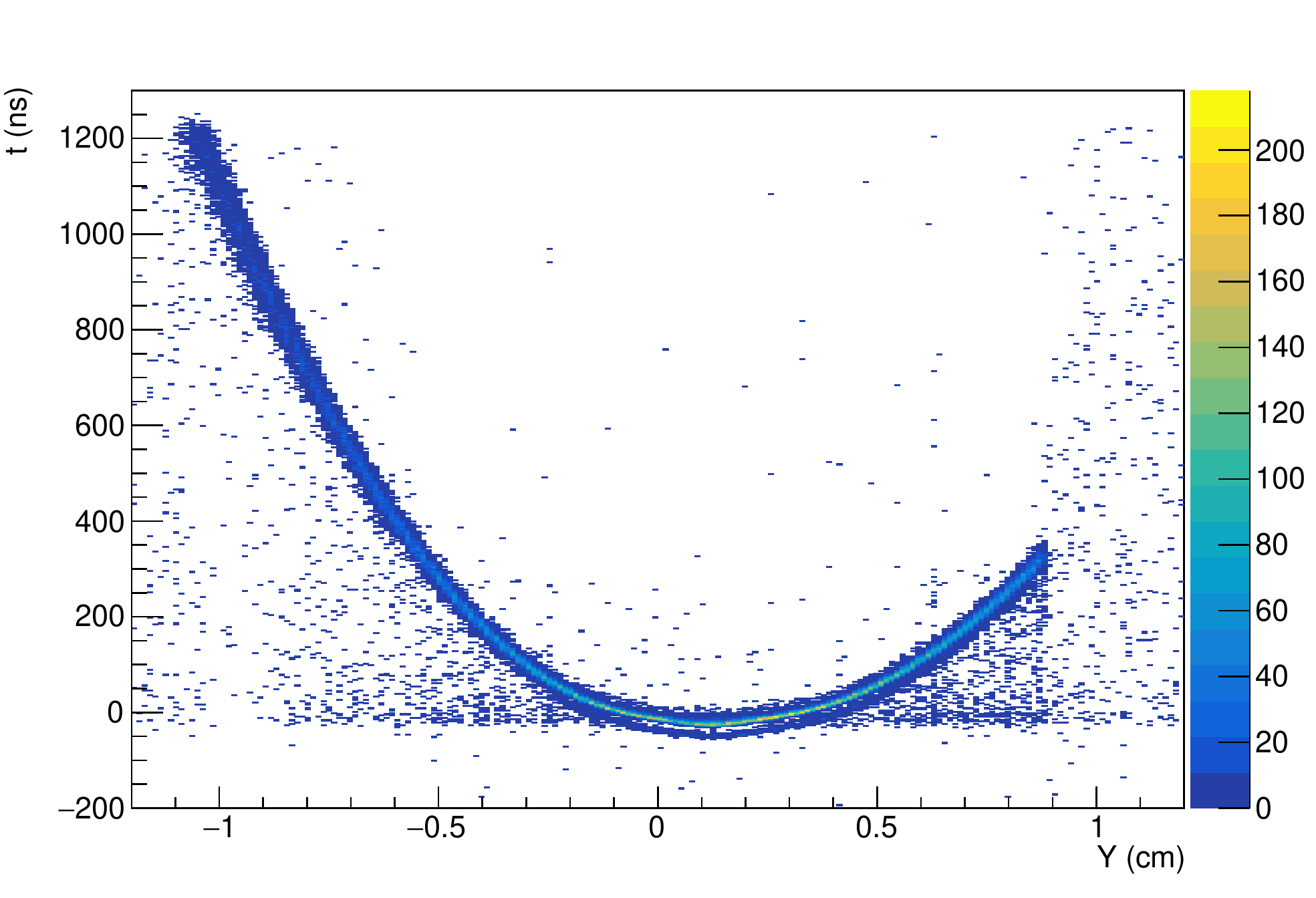}
    \caption{Measured drift time versus Y position of the reconstructed particle trajectory
with the 20\,mm diameter straw-tube prototype with a large wire eccentricity (2.05\,mm).
The Y axis is vertical, perpendicular to the wire axis X and to the particle beam axis Z.}
    \label{fig:straw-V-shape-with-offset}
\end{figure}

\begin{figure}
    \centering
    \includegraphics[width=0.99\linewidth]{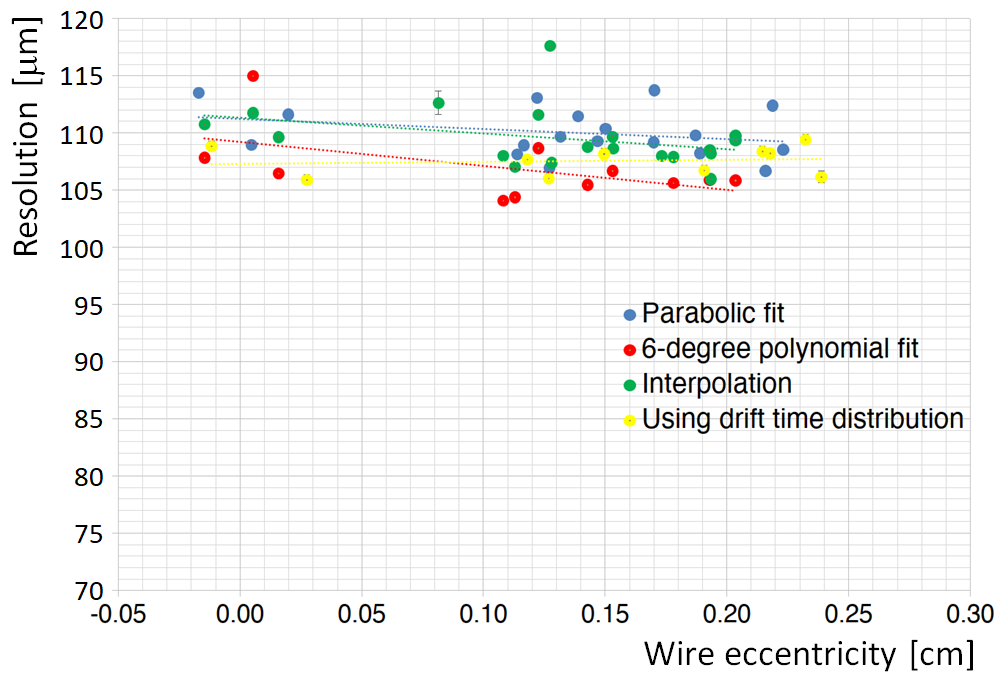}
    \caption{Measured spatial resolution for the 20\,mm diameter straw tube prototype as a 
function of the artificially induced wire eccentricity. Four different analysis methods of the spatial resolution are compared (see CERN-THESIS-2020-218). The lines are the results of linear fits.}
    \label{fig:HS_spatial}
\end{figure}

The main element of the HSDS is the straw tracker designed to accurately reconstruct the decay vertex, the mass, and the impact parameter of the hidden particle trajectory with respect to the proton interaction region. The precision of the extrapolated tracks must be well matched with the segmentation of the downstream timing detector. 

The straw tracker design consists of two tracking telescopes located in the vacuum vessel, upstream and downstream of the magnet, and each composed of two tracking stations. The four stations are identical with a nominal acceptance of 5\,m in width and 10\,m in height, and are based on 20\,mm diameter ultralight drift tubes inspired by the NA62 design~\cite{NA62strawtracker}. The cathode is constructed from $36\,\upmu$m thick PET film coated with 50\,nm Cu and 20\,nm Au. The anode is made of an Au-plated tungsten wire. As a consequence of the vertical bending of the spectrometer magnet, the straw tubes are oriented horizontally and have a length of 5\,m.  
Background simulations indicate that, for the chosen straw dimensions, the expected straw hit rates will remain everywhere below 10\,kHz. Each station contains four views, in a Y-U-V-Y arrangement, where U and V are stereo views
with straws rotated by a small angle $\pm\theta_{\rm stereo}$ around the Z-axis with respect to
the Y-measuring straws. In the baseline, $\theta_{\rm stereo} = 5^\circ$. In total, the four views consist of about 4000 straw tubes which, together with services, are mounted on a frame. The frame, hung from a cover plate, is lowered into the vacuum vessel through openings in the roof of the vessel (Fig.~\ref{fig:HS_SpectrometerSection}). Straw tube elongation and relaxation effects present a serious challenge to the mechanics, which, if neglected, would result in an evolving and possibly excessive sagging of the straws. Three mechanical designs have been developed. The first one includes a constant-force spring to maintain the wire under the desired tension while accommodating the evolving straw elongation of several centimeters with the help of an expandable frame. The second scheme utilizes a straw suspension mechanism based on carbon fibres. The third one is inspired by the self-supporting design of the PANDA straw tracker~\cite{7721e0933a6f4b8f8e5492f4632cc250}.


A 2\m long prototype straw was manufactured, with a 30\,\textmu m diameter anode, and its performance was characterised in a test run with beam in the CERN SPS North area 
as a function of the wire eccentricity at nominal conditions (2.2~kV, $\sim 1.05$\,bar pressure, 
70\% Ar / 30\% CO$_2$). The tracks were externally measured by a silicon-strip telescope made of
8 single-sided sensors~\cite{Milano-telescope} and a 2.5\,cm thick plastic scintillator for defining the start time. Fig.~\ref{fig:straw-V-shape-with-offset} shows the measured drift time versus the vertical position of the extrapolated track at the straw for a 2.05\,mm artificially induced wire eccentricity.
 As shown in Fig.~\ref{fig:HS_spatial}, a straw hit resolution of <120\,\textmu m is achievable with high hit efficiency over most of the straw diameter, independently of the wire eccentricity.

\subsection{HSDS timing detector}

\begin{figure}
\centering 
\includegraphics[width=0.95\linewidth]{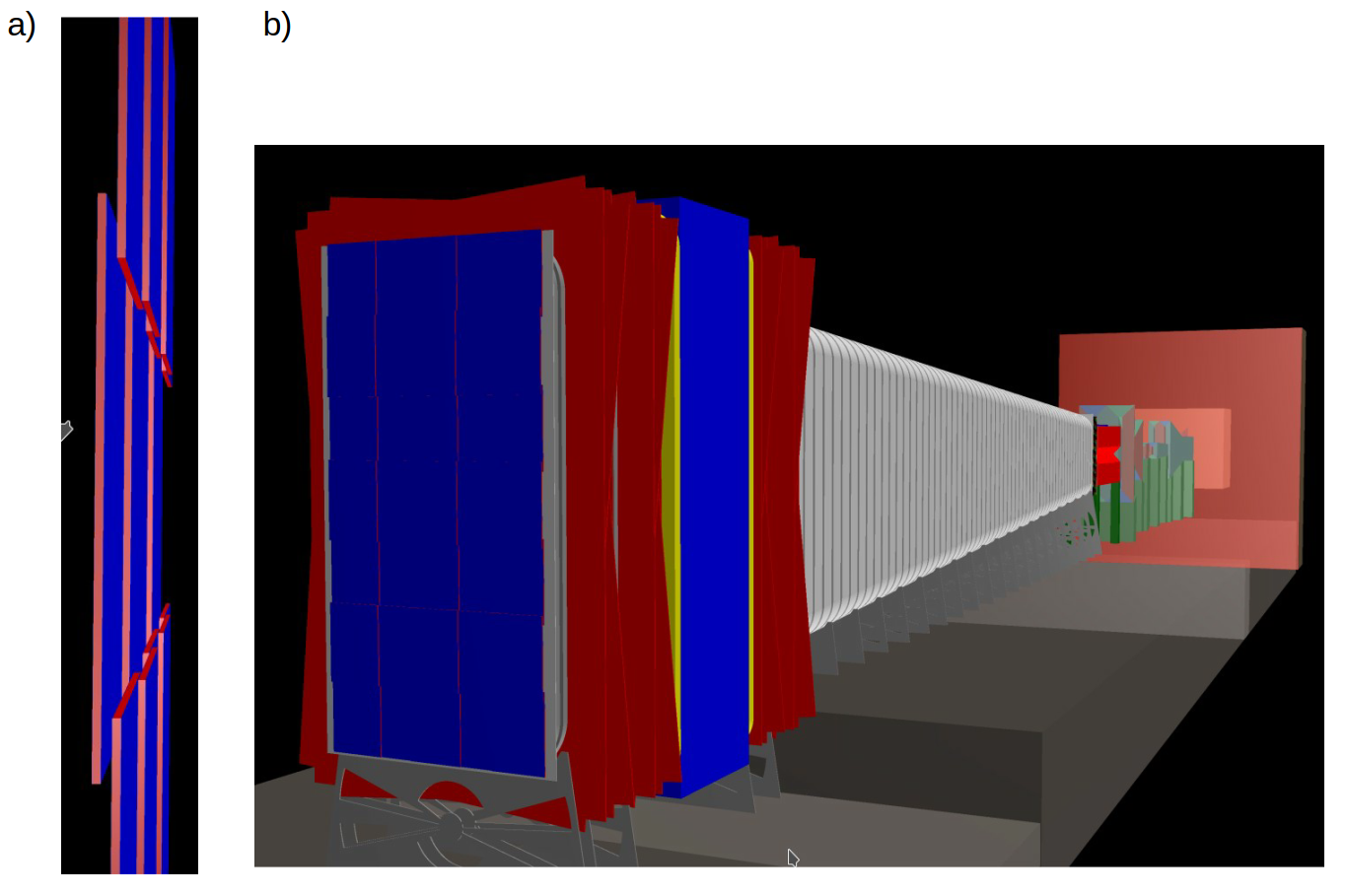}
    \caption{ Side view (a) and downstream view (b) of the HSDS timing detector from the geometry implemented in the GEANT4-based~\cite{GEANT4} physics simulation.}
\label{fig:TD_MRPC}
\end{figure}

The timing detector covers the $50$\,m$^2$ physics aperture of the vacuum vessel end-cap window (Fig.~\ref{fig:TD_MRPC}). Its purpose is to provide information on the coincidence of the charged particles originating from a decay candidate. At the expected rate of fully reconstructed residual background muons of $\sim25\,\khz$, a hit time resolution of $<100\,\ps$ is required in order to provide sufficient rejection of combinatorial events. Two technologies are currently being considered: scintillating bars and MRPCs.

\subsubsection{Scintillator-based option}

The scintillator option uses EJ200 plastic bars read out by large-area SiPMs.  The material for the scintillator plastic was chosen by the requirement on the timing resolution. EJ200
 is found to have the right combination of light output, attenuation length (3.8\,m) and fast timing (rise time of 0.9\,ns and decay time of 2.1\,ns). The emission spectrum peaks at 425\,nm, perfectly
 matching the SiPMs spectral response. The number of photons generated by a minimum-ionising particle
 crossing 1\,cm of scintillator is $\mathcal O(10^4)$. The bars are wrapped in an aluminum foil, and a black plastic
 stretch film on top, to ensure opacity.

The $5\m\times10\m$ aperture is built from three columns, each with 182 rows of plastic bars with dimensions 168\,cm $\times$\,6\,cm\,$\times$\,1\,cm. A 0.5\,\cm overlap between columns and a 1\,cm overlap between bars in the same column provide means of cross-calibration during time alignment. Each bar is read out on both sides by a matrix of eight 6\,mm\,$\times$\,6\,mm SiPMs. The signals from the eight SiPMs are summed to form a single channel, hence making up 1092 channels which are subsequently digitized by a DAQ module based on the SAMPIC ASIC. 

A 22-bar (44 channels) prototype array with 1.68\,m long bars has been successfully operated at the CERN PS test beam~\cite{Betancourt:2020gyr}. The resolution was demonstrated to be $\sim 80$\,ps over the whole 2.1\,m$^2$ area of the prototype. 

\subsubsection{MRPC-based option}
\label{subsec:TDMRPC}

\begin{figure*}
\centering 
\includegraphics[width=0.8\linewidth]{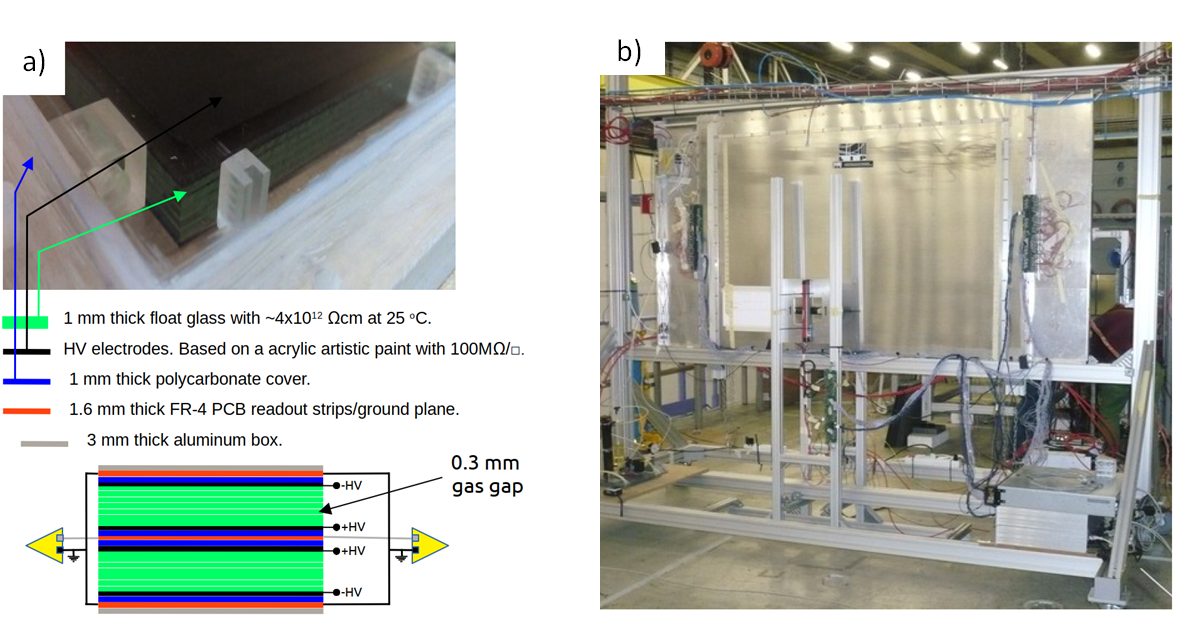}
\caption{a) Detailed photograph of one of the MRPCs together with a sketch of the internal structure of the module. b) Experimental setup at the T9 beam line (CERN East Area), where the MRPC was exposed to negative pions of $8$\,GeV/c.}
\label{fig:setup}
\end{figure*}

In the MRPC option, the end-cap window of the vacuum vessel is covered by 35 modules, each with two stacked MRPC chambers, arranged in a matrix $7$ (vertically) $\times$ $5$ (horizontally). The active areas of the modules are overlapped by $10$\,cm (vertically and horizontally) in order to create a layer without dead regions.

The MRPCs use a novel concept ~\cite{Blanco_2015, Lopes_2019} in the construction of the chambers, in which the glass stack and High-Voltage (HV) electrodes are confined within a permanently sealed plastic box. PMMA is used for the frame and polycarbonate (PC) is used for the covers. The box is equipped  only with gas and HV feed-throughs (Fig.~\ref{fig:setup}a). This facilitates construction and allows operation with a low gas flow of a few cm$^3$/(min$\times$m$^2$).

Each MRPC chamber has six gas gaps defined by seven $1$\,mm thick float glass electrodes of about $1550~\times~1250$\,mm$^2$ separated by $0.3$\,mm nylon mono-filaments. The glass has a bulk resistivity of $\approx 4\times10^{12}$\,$\upOmega$cm at $25$\,$^\circ$C. The HV electrodes are made up of a semi-conductive layer based on an acrylic paint with sheet resistivity of around 100\,M$\upOmega/\Box$ that is applied to the outer surface of the outermost glasses with airbrush techniques.

The two MRPCs chambers are read out in parallel by a readout strip plane which is based on a $1.6$\,mm Flame Retardant\,4 (FR4) printed circuit board located between the two chambers. The readout plane is equipped with 41 copper strips with $29$\,mm with, $30$\,mm pitch and $1600$\,mm length. Two ground planes, located on the top and bottom of the dual MRPC stack complete the readout planes. The complete structure is housed in an aluminum box that provides the necessary electromagnetic insulation and mechanical \linebreak rigidity. A schematic of the inner structure of the module is shown in Fig.~\ref{fig:setup}a. 
The module operates in a open gas loop with a mixture of $97.5$\%  C$_{2}$H$_{2}$F$_{4}$ and $2.5$\% SF$_{6}$ at a pressure a few millibars below the atmospheric pressure. In this way, the width of the gaps is correctly defined with the help of the compression exerted by  atmospheric pressure. 

The MRPC signals from both sides of each strip are fed to fast front-end electronics~\cite{HADES_FEE} capable of measuring time and charge in a single channel. The resulting signals are read out by the ``TRB3'' board~\cite{TRB3} equipped with $128$ multi-hit TDC (TDC-in-FPGA technology) channels with a time precision better than $20$\,ps RMS.

\begin{figure}
\centering 
\includegraphics[width=0.99\linewidth]{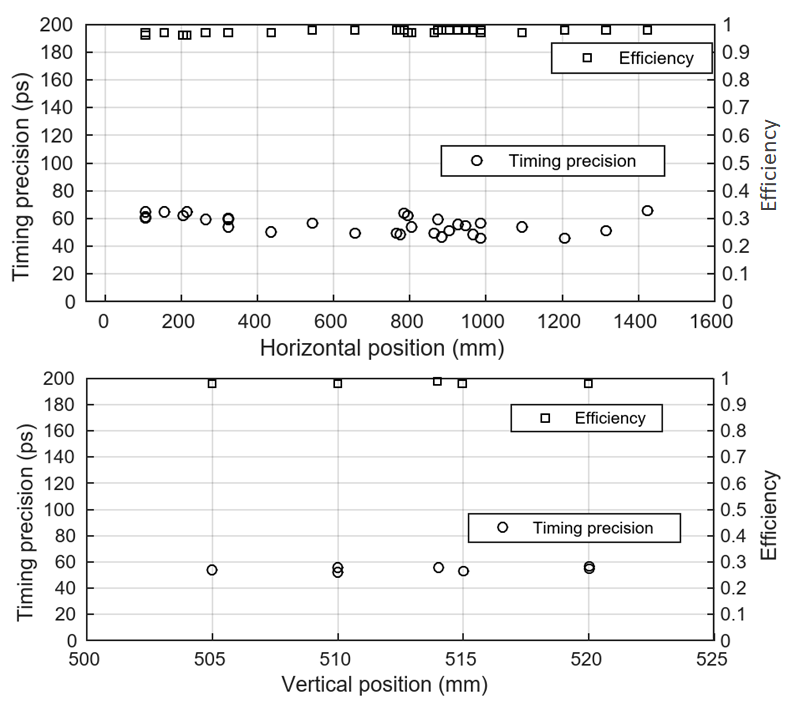}
    \caption{Time resolution and efficiency as a function of the horizontal and vertical position on the MRPC prototype.}
\label{fig:T_EvsXY}
\end{figure}

A complete prototype was exposed to $8$\,GeV negative pions in the T9 beam line (CERN East Area) (Fig.\ref{fig:setup}b). Time resolution and detection efficiency were measured at various positions over the active area. Fig.~\ref{fig:T_EvsXY} shows the time resolution and efficiency as a function of the horizontal position (for a vertical position of $500$\,mm) and vertical position (for a horizontal position of $725$\,mm) respectively without noticeable dependence on the position. The measurements demonstrate a time resolution of $54$\,ps and an efficiency of 98\%. More details on the beam time results can be found in~\cite{SHIPRPC}.

\subsection{HSDS electromagnetic calorimeter}

\begin{figure*}[ht]
    \centering
    \includegraphics[width=0.7\linewidth]{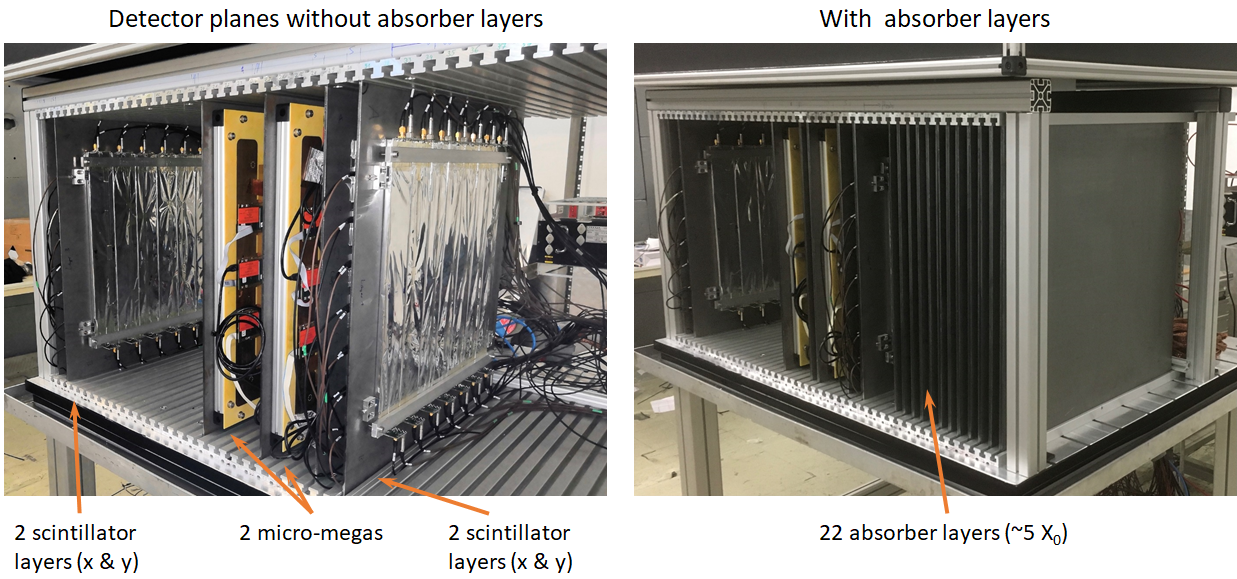}
    \caption{Test beam setup for the electromagnetic calorimeter measurements of the shower profiles.}
    \label{fig:ecal_tb}
\end{figure*}


\begin{figure*}[ht]
    \centering
    \includegraphics[width=0.8\linewidth]{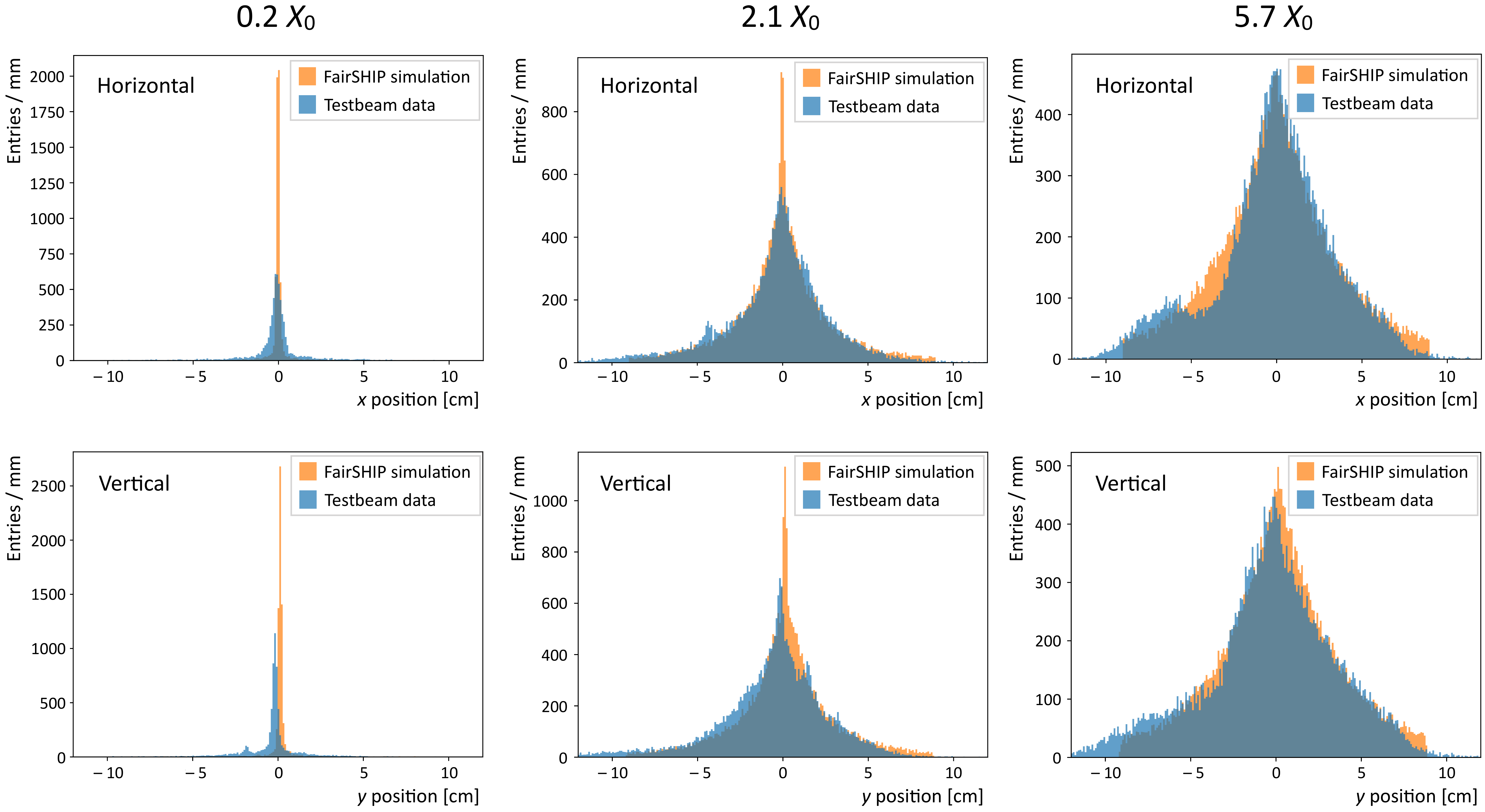}
    \caption{Measured and simulated transverse electromagnetic-shower distribution with 5\,GeV electrons from the CERN PS at different calorimeter depths. The small discrepancies between simulation and data are mostly due to known issues of the prototype read-out.}
    \label{fig:ecal_meas}
\end{figure*}

Apart from providing electron and photon identification and discriminating between hadrons and muons, the electromagnetic calorimeter should be capable of measuring the shower angle with a resolution of $\sim$5\,mrad to reconstruct two-photon final states. In the baseline configuration, SHiP is not \linebreak equipped with a hadron calorimeter. The longitudinal segmentation of the  electromagnetic calorimeter, with the shower energy being  sampled every 0.5\,$X_0$, provides sufficient electron/hadron separation.

The electromagnetic calorimeter is based on the ``SplitCal'' concept~\cite{Bonivento:2018eqn}. It consists of a longitudinally segmented lead sampling calorimeter with a total sampling depth of 20\,$X_0$. The lead absorber plates are 0.5\,$X_0$ thick, i.e. 0.28\,cm, thus leaving space for 40 sampling layers.
Most sampling layers are equipped with scintillating plastic bars read out by WLS fibres with a relatively coarse spatial segmentation. The scintillator planes are 0.56\,cm thick. 

Three sampling layers, each with a thickness of 1.12\,\cm, are equipped with high resolution detectors providing a spatial segmentation of $\sim$200\,$\upmu$m. They are located at a depth of 3\,$X_0$ and at the shower maximum in order to accurately measure the barycentres of the transverse shower profile. The shower direction is determined from the three measurements of the barycentres.
For the high-resolution layers, it is foreseen to use  micro-pattern detectors, such as micro-megas.
 
To increase the lever arm for the angular measurement, the calorimeter is mechanically split in two parts in the longitudinal direction with an air gap  of 1\,m between the first 3\,$X_0$ and the remaining 17\,$X_0$. With a few mm transverse shower-position resolution in the high-precision layers, the target angular resolution is of the order of a few mrad. 
 
The principal challenge in achieving a good angular resolution, along with high efficiency for the photon reconstruction, is the presence of shower satellites due to long tails in the transverse shower shape.  The shower profiles were measured with a prototype at an electron beam test at CERN in order to tune the simulation and optimise the layout.
The test setup is shown in Fig.~\ref{fig:ecal_tb}, while the experimental result comparing data and simulation at different shower depths is shown in Fig.~\ref{fig:ecal_meas}.

\subsection{HSDS muon system}


\begin{figure}
    \centering
    \includegraphics[width=0.99\linewidth]{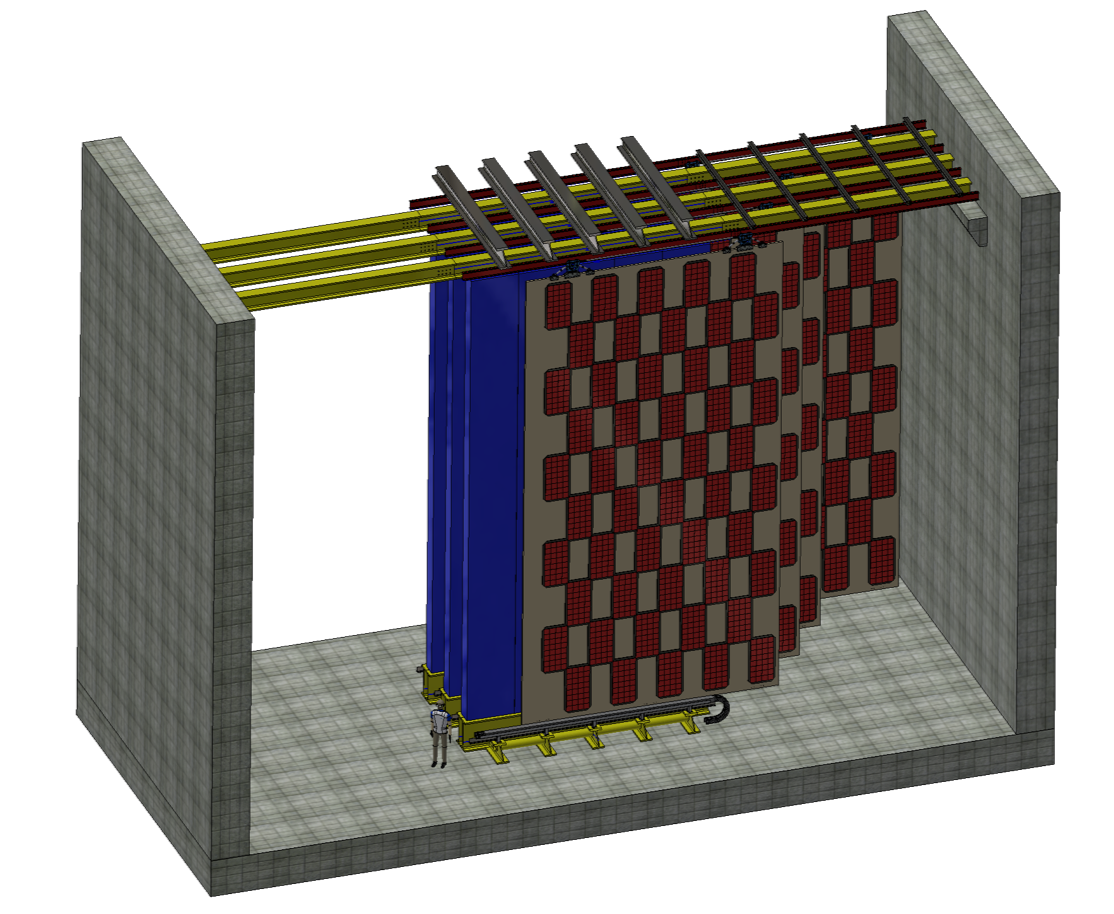}
    \caption{Conceptual layout of the HSDS muon system made of four stations.}
    \label{fig:HS_MuonSystem}
\end{figure}

 The muon system provides identification of muons with an efficiency of $> 95\%$ in the momentum range of \linebreak $\sim 5-100$\,\gevc with a mis-identification rate of $\sim 1-2\,\%$.
 
As shown in Fig.~\ref{fig:HS_MuonSystem}, the muon system is comprised of four stations of active layers interleaved by the three muon filters.
The four stations are 6\,m wide and 12\,m high.  
The amount of material of the calorimeter system corresponds to 6.7 interaction lengths ($\lambda_I$). 
The muon filters are 60\,cm thick iron walls, corresponding to 3.4\,$\lambda_I$ each.
A muon with normal incidence must have an energy of at least of 2.6\gev to reach the first muon station 
and at least 5.3\gev to reach the last muon station.
The multiple scattering of  muons in the material of the calorimeter and the muon filters drives the 
granularity of the system. Simulation studies show that a readout granularity of $\sim$10\,cm in the transverse directions
is adequate for the momentum range of interest.

The rate expected in the muon detector is between 1 and 6\,MHz, depending on the station, mostly caused by the beam-induced muon background. 
Simulation studies show that the rate is dominated by very low momentum \linebreak ($<$100\,\mevc) electrons, positrons and photons
produced by inelastic and electromagnetic interactions of muons with the material of the detector.
Most of the hits are concentrated in the first muon station. The second and third stations see very low rates while the fourth station
sees an almost uniform illumination originating from very low momentum particles surrounding the system.
To shield the last muon station against hits arising from muon interactions with the surrounding material, including the cavern walls, a thin ($\sim$10\,cm) layer of iron is located downstream of the last station.
The hit rate from real muons is subdominant and does not exceed $\sim$300\,kHz, being concentrated on the sides of stations.

The detectors of the muon system cover the total surface of $\sim288 $\,m$^2$. Scintillating bars and scintillating tiles with direct SiPM readout were investigated as options for SHiP. All details of these works, including performance results from beam tests, can be found in~\cite{Baldini_2017, Balla_2022}.

\section{Common detector electronics and online system}
\label{sec:ElectronicsOnline}

The design of the SHiP front-end electronics and readout system is characterised by a relatively small data throughput, no radiation to the electronics, and mostly trivial powering and cooling. The complexity lies in the collection of data from a relatively high number of channels spread out over a very large detector, and in the event building with a very wide range of times-of-flight.

\begin{figure*}
    \centering
    \includegraphics[width=0.99\linewidth]{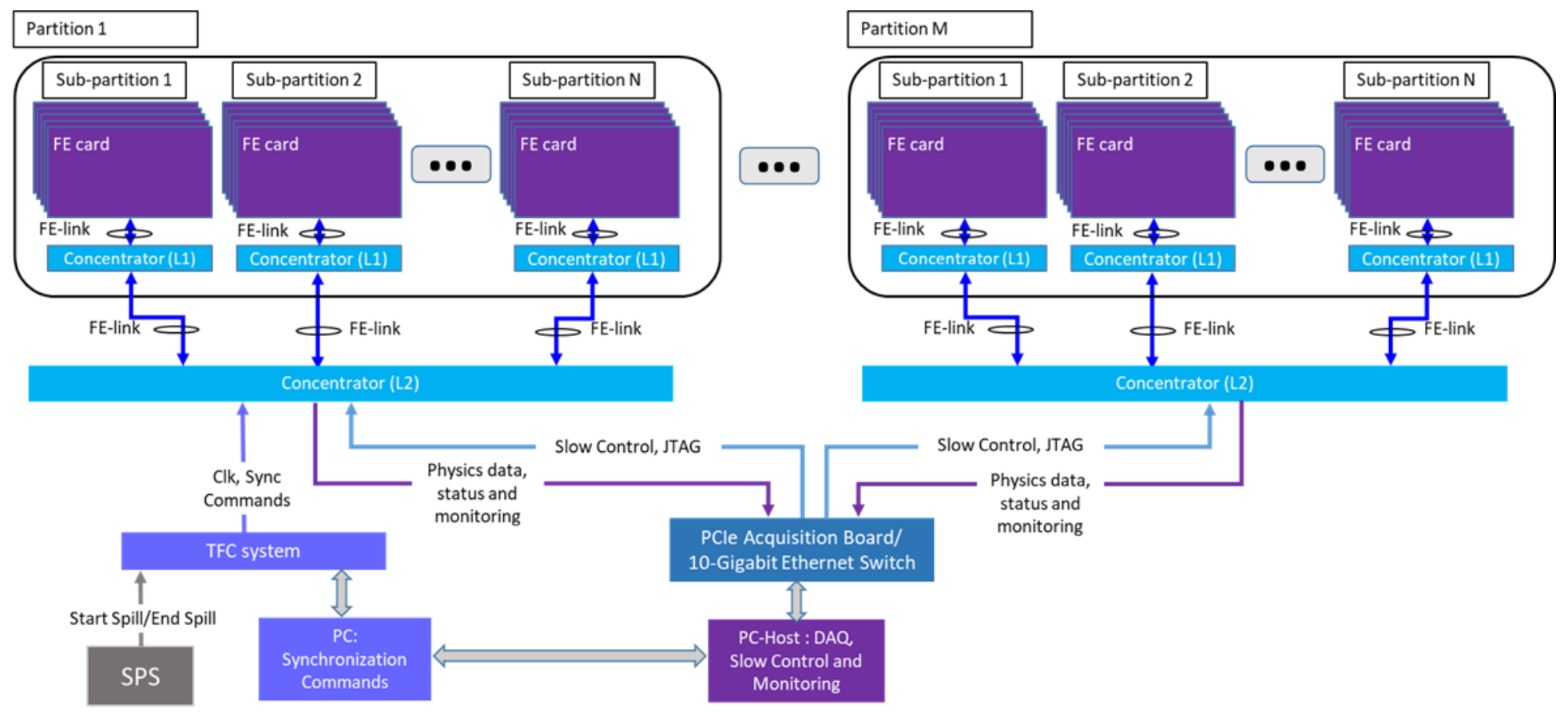}
    \caption{Global scheme of the SHiP electronics and readout system.}
 \label{fig:Ele_diagram}
\end{figure*}

\begin{figure}
    \centering
    \includegraphics[width=0.99\linewidth]{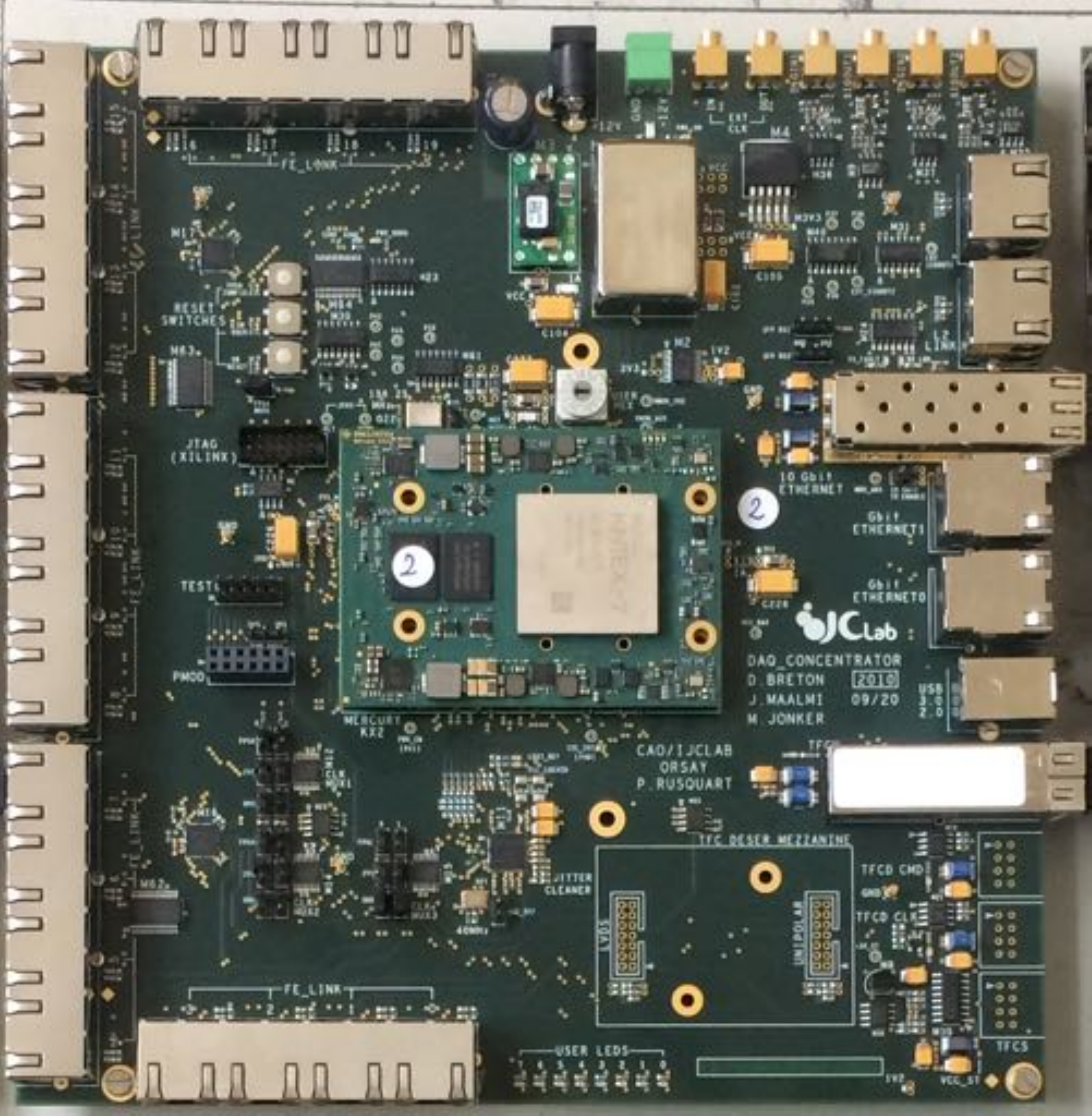}
    \caption{Prototype of the SHiP FE concentrator board.}
    \label{fig:Ele_concentrator}
\end{figure}

Fig.~\ref{fig:Ele_diagram} shows an overview of the electronics and readout system. The system has two main subsystems: the control distribution, data transport and concentration (CTC) system; and the timing and fast control (TFC) system.  Downstream of the front-end (FE) electronics, the system is composed of cascaded FE concentrators which fan-in and fan-out the CTC and the TFC networks. The FE links are based on 4 LVDS copper pairs carrying physics data at 400 Mbits/s, 40 MHz clock, fast commands and slow control at 40 Mbits/s, and status monitoring at 40 Mbits/s, respectively. Fig.~\ref{fig:Ele_concentrator} shows a photo of a prototype of the FE concentrator.
Downstream of the FE concentrator chain of each subsystem (called partition), the last concentrator is interfaced with a front-end host (FEH) computer for data readout, slow control and monitoring, and with the TFC master for the clock and synchronous commands, as shown in Fig.~\ref{fig:Ele_diagram}. The design strategy is to base the system as much as possible on FPGAs, including the FE electronics. 

The architecture does not comprise a hardware trigger. The FEH computers format the data and forward them to the event filter farm (EFF). For every SPS cycle, a computer in the EFF is assigned to collect the partition data, to extract the physics events candidates and to build the events. The EFF performs reconstruction and event triggering after the final event building with data from complete SPS cycles. The FEH and EFF computers are based on commodity PCs. 

With the baseline detector, it is estimated that about 300 concentrator boards will be required together with a total of 25 DAQ links, 12 FEH and 42 EFF computers. 
\section{Simulation and reconstruction}
\label{sec:simulation}

The SHiP experiment relies on an accurate simulation of the background components, in particular related to combinatorial events from muons, and muon and neutrino deep inelastic scattering (DIS) in the material of the detector. The design of the muon shield also relies on an accurate knowledge of the muon spectrum. The SHiP software framework for simulation, reconstruction, and analysis is based on the FairRoot package~\cite{FairRoot} and is called FairShip. The FairShip code is largely a specialisation of the generic C++ base classes provided by FairRoot, mainly profiting from the methods to describe the detector geometry, implement detector classes, and performing simulation and reconstruction. The framework incorporates \textsc{GEANT4}~\cite{GEANT4} to trace particles through the target and the experimental setup, \textsc{PYTHIA8}~\cite{pythia8} for the primary proton fixed-target interaction, \textsc{PYTHIA6}~\cite{Sjostrand:2006za} for muon deep inelastic scattering, and the \textsc{GENIE}~\cite{Andreopoulos:2009rq} MC generator for interactions of neutrinos.  In addition some specific auxiliary libraries needed by SHiP are included, like GENFIT~\cite{genfit2015} for reconstruction of tracks. The steering of the simulation flow, and the main parts of the reconstruction and analysis are based on Python classes and functions.

For background studies and detector optimisation, a total of $6.5\times 10^{10}$~protons on target have been simulated with an energy cut of 10\,\gev for transporting particles after the hadron absorber. This simulation was run with strongly enhanced muon production from QED processes, such as resonance decays and gamma conversion, such that for the studies of muon-induced backgrounds, the sample corresponds to $6.5\times 10^{12}$~protons on target. In addition, a total of $1.8\times 10^9$~protons on target have been simulated with an energy cut of 1\gev. Heavy flavour production is both a source of signal and background. Dedicated samples  of charm and beauty hadrons corresponding to about $10^{11}$~protons on target have been produced. These simulation samples give sufficient statistics after the muon shield for the background determination to be extrapolated to full run of SHiP with $2\times 10^{20}$~protons on target with good statistical accuracy and such that any rare contribution to the muon flux is subdominant. 

In order to produce a background sample of muon DIS events that is equivalent to what is expected for the full run of SHiP, the muon samples from the simulations above were used to produce DIS events with the cross-section boosted such that every muon interacts according to the material distribution of the experimental setup. 

For neutrino DIS, the neutrino spectra from the simulated minimum bias, and charm and beauty samples were used to produce a sample of neutrino interactions in the material of the detector with the help of the \textsc{GENIE} generator that is equivalent to seven times the full run of SHiP, again by making every neutrino interact according to the material distribution of the experimental setup.

The validity of the FairShip prediction of the particle fluxes has been verified by comparing to the data from the CHARM beam-dump experiment at CERN~\cite{Dorenbosch:164101}(see PhD thesis F. Bergsma, University of Amsterdam, 1990). 
The most realistic cross-check of FairShip has been performed in summer 2018 in a dedicated experiment at the CERN SPS~\cite{Ahdida:2020doo}. It has directly measured the rate and momentum of muons produced by $400\gevc$ protons dumped on a replica of the SHiP target, and found a very good agreement between the prediction by the simulation and the measured spectrum.

The production and decays of the various HS particles have been implemented in FairShip. \textsc{PYTHIA8} is predominantly used to generate the different signal processes. For particles, and background, produced from the decays of charm and beauty hadrons, the effect of cascade production of charm and beauty from secondary hadrons is accounted for. For decays to hadronic final states, fragmentation is handled by \textsc{PYTHIA8}.


\section{Expected detector performance}

The SHiP detector performance has been studied in detail with the help of the full MC FairShip framework supported by the measurements done in test beam on the detector prototypes.
The physics performance of the experiment is anchored in a highly efficient background suppression, provided by the target design, hadron stopper and the muon shield. The background suppression in the HS decay search is further guaranteed by the vacuum volume and the background taggers. The overall detector concept provides sensitivity to as many decay modes as possible to ensure model-independent searches. 

In addition to improving present constraints on many models by several orders of magnitude, SHiP's decay spectrometer allows distinguishing between different models, and, in a large part of the parameter space, measure parameters that are relevant for model building and cosmology.  These features make the SHiP experiment a unique discovery tool for HS particles.  Moreover, together with the direct search for LDM, and neutrino physics, SHiP represents a wide scope general-purpose beam-dump experiment. 



\subsection{SND performance}

\label{sec:neutrino_physics}
The nuclear emulsion technology combined with the information provided by the SND muon identification system makes it possible to identify the three different neutrino \linebreak flavours in the SND detector.
The neutrino flavour is determined through the flavour of the primary charged lepton produced in neutrino charged-current interactions. The muon identification is also used to distinguish between muons and hadrons produced in the $\tau$ decay and, therefore, to identify the $\tau$ decay channel. In addition,
tracking in the SND magnetic spectrometer will allow for the first time to distinguish between $\nu_\tau$ and $\overline{\nu}_\tau$ by measuring
the charge of $\tau$ decay products. The charge sign of hadrons and muons is measured by the CES, the downstream tracker, and by the muon identification system. 

The neutrino fluxes produced in the beam dump have been estimated with FairShip, including the contribution \linebreak from cascade production in the target. 
The number of charged-current deep inelastic scattering (CC DIS) interactions in the neutrino target is evaluated by convoluting the generated neutrino spectrum with the cross-section provided by the \textsc{GENIE}~\cite{Andreopoulos:2009rq} Monte Carlo generator. The expected number of CC DIS in the target of the SND detector is reported in the first column of Table~\ref{tab:neu_yield}. 
\begin{table}
\begin{center}
\begin{tabular}{c   c c }
&   CC DIS  & CC DIS  \\
&  interactions   & w. charm prod.\\
 \hline\hline
 $N_{\nu_e}$ & $8.6 \times 10^{5}$ & 5.1 $\times$10$^4$ \\
 $N_{\nu_\mu}$ & $2.4 \times 10^{6}$ &  1.1 $\times$10$^5$  \\
 $N_{\nu_\tau}$ & $2.8 \times 10^{4}$  & 1.5 $\times$10$^3$\\
 $N_{\overline{\nu}_e}$ & $1.9 \times 10^{5}$ & 9.8 $\times$10$^3$ \\
 $N_{\overline{\nu}_\mu}$ & $5.5 \times 10^{5}$ & 2.2 $\times$10$^4$ \\
 $N_{\overline{\nu}_\tau}$  & $1.9 \times 10^{4}$  & 1.1 $\times$10$^3$\\
 \hline
\end{tabular}
  \caption{ \label{tab:neu_yield} Expected CC DIS interactions in the SND assuming $2\times10^{20}$ protons on target.}
\end{center}
\end{table}


%
 %

With $2\times 10^{20}$ protons on target, more than $\sim$2$\times10^5$ \linebreak neutrino-induced charmed hadrons are expected, as reported in the second column of Table~\ref{tab:neu_yield}. The total charm yield exceeds the samples available in previous experiments by more than one order of magnitude.
No charm candidate from electron neutrino interactions
was ever reported by any previous experiment. Consequently, all the studies on charm physics performed with neutrino interactions will be improved, and some channels inaccessible in the past
will be explored. This includes the double charm production cross-section~\cite{KayisTopaksu:2002fv, Abt:2007zg} and the search for pentaquarks with charm quark content~\cite{DeLellis:2007kz}. Charmed hadrons produced in neutrino interactions are also important to investigate the strange-quark content of the nucleon. The samples available at SHiP will also allow to significantly constrain the $\nu_\tau$ magnetic moment and test lepton flavour violation in the neutrino sector.

The SND can also probe existence of LDM ($\chi$) by detecting the electromagnetic showers initiated by the recoil electrons coming from elastic scattering of LDM in the SND. The SND ECC bricks, interleaved with the SND target \linebreak tracker planes, act as sampling calorimeters with five active layers per radiation length, $X_{0}$, and a total depth of $10\,X_{0}$. The configuration allows reconstructing a sufficient portion of the shower produced by the recoil electron to determine the particle angle and energy. In addition, the micro-metric accuracy of the nuclear emulsions provides crucial  topological discrimination of LDM interactions against neutrino-induced background events.

Neutrino events with only one reconstructed outgoing electron at the primary vertex constitute background in the LDM searches, mimicking the signature $\chi e^{-}\to \chi e^{-}$. The \textsc{GENIE} Monte Carlo generator, interfaced with FairShip, has been employed for a full simulation to provide an estimate of the expected background for $2\times 10^{20}$ protons on target. After imposing a selection optimised for the signal, the residual neutrino background amounts to 230 events.
All results of this study and the SHiP sensitivity to LDM are reported in~\cite{SHiP:2020noy}. 

\begin{figure}
\centering
\includegraphics[width=0.99\linewidth]{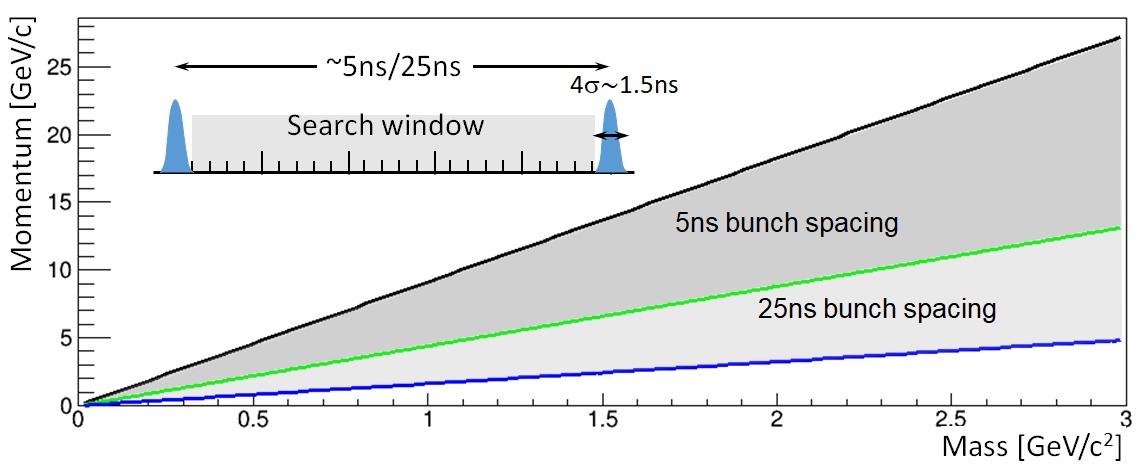}
\caption {Mass-momentum region in which it is possible to discriminate an LDM scattering signature from a neutrino interaction using slow extraction of bunched beam in the SPS and a time-of-flight measurement. The figures assumes a bunch structure with $5\ns$ respectively $25\ns$ bunch spacing and $1.5\ns$ $(4\sigma)$ bunch length, and a distance of $40\unit{m}$ between the target and the detector.}
\label{fig:BunchedBeam}
\end{figure}

In order to further reduce the background from neutrino events, and consolidate a possible LDM signal, it has been envisaged to alternatively operate SHiP with slowly extracted spills of bunched beam instead of uniformly de-bunched beam. Using the timing capability of the SND target tracker allows rejecting the neutrino background based on the difference in the time of flight. Fig.~\ref{fig:BunchedBeam} shows the region of discrimination assuming the 4$\sigma$ SPS bunch length of 1.5\,ns and a bunch spacing of 25\,ns and 5\,ns, and
40\,m distance between the beam-dump target and the detector. Currently, this mode of operation would only be used in order to consolidate a signal. Further studies are needed to determine if this mode of operation could also work for the HS decay search.

\subsection{HSDS performance}

\begin{figure*}
    \centering
    \includegraphics[width=0.40\linewidth]{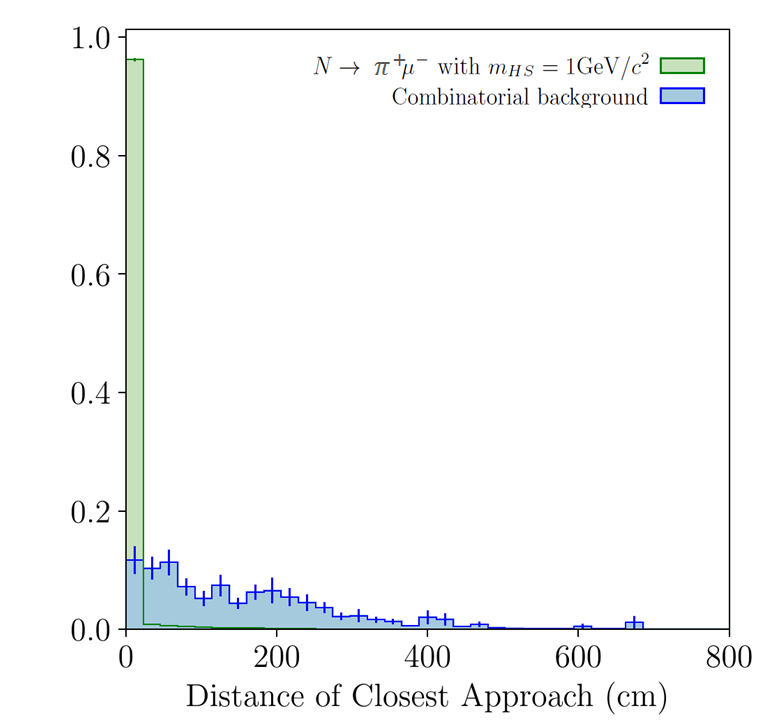}
    \includegraphics[width=0.40\linewidth]{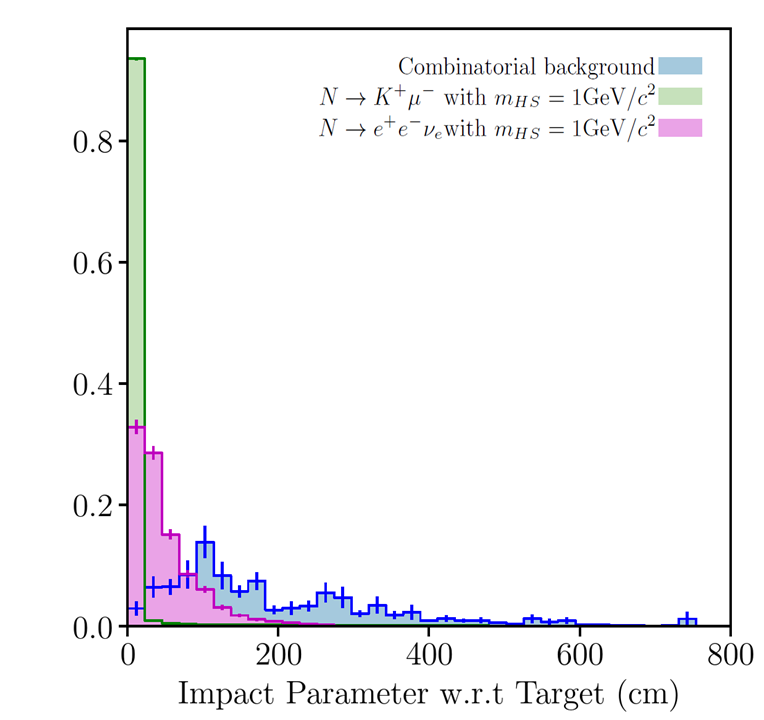}    
    \includegraphics[width=0.40\linewidth]{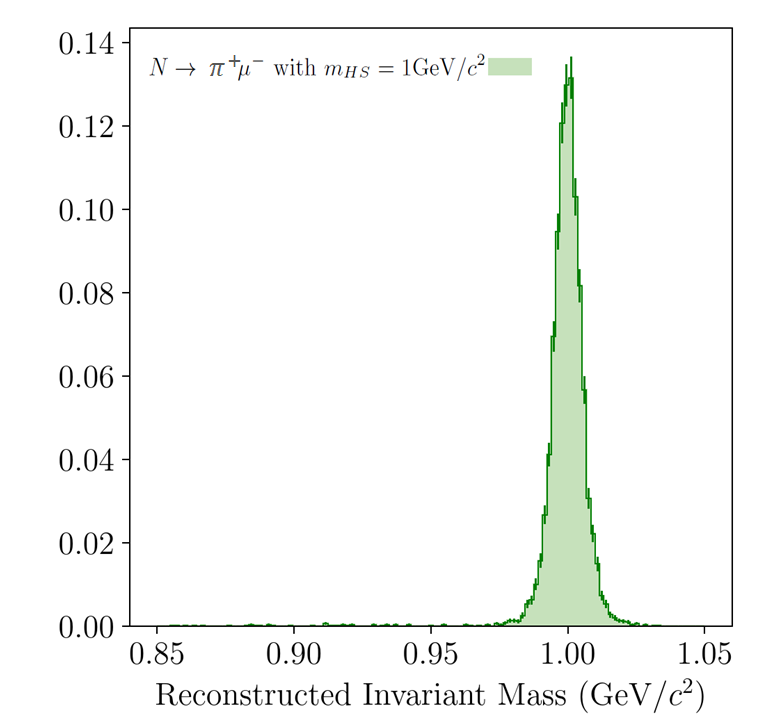}
    \includegraphics[width=0.40\linewidth]{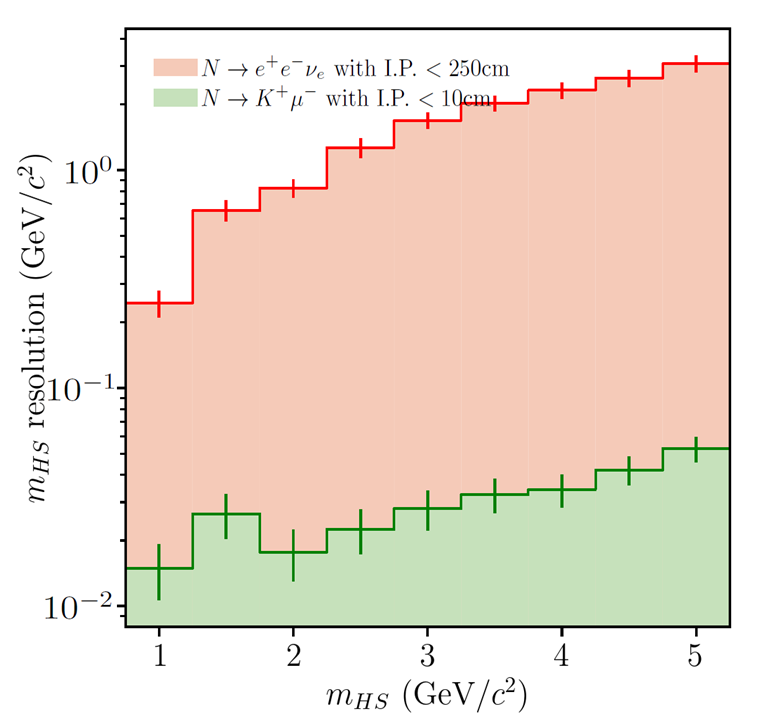}
    \caption{Examples of detector performance showing key reconstructed observables for an HNL candidate of $m_{HS}=1$\,GeV/c$^2$ decaying to fully reconstructed modes and to partially reconstructed modes with neutrinos.  For the distance of closest approach (top-left) and impact parameter (IP) with respect to the proton interaction region (top-right), both the signal (green and pink) and combinatorial background (blue) distributions are shown. Details on the combinatorial background can be found in Section~\ref{sec:bkg_rejection}. All distributions are normalised to unit area. Bottom-left shows the reconstructed invariant mass and bottom-right the mass resolution as a function of the HNL mass.}
    \label{fig:HS_Performance_FullyReconstructed}
\end{figure*}

The broad range of signals to which the SHiP experiment is sensitive can be classified into two main categories: fully and partially reconstructed decays. The former category refers to decays where there are at least two charged particles and no invisible final state particles, examples are DP$\to \mu^\pm\mu^\mp$ and HNL$\to \mu^\pm\pi^\mp$. 
The latter category refers to decays with at least two charged particles and at least one invisible particle in the final state, e.g. HNL$\to \mu^\pm\mu^\mp\nu$.
In all cases, the experimental signature consists of an isolated vertex with a total momentum vector that extrapolates accurately back to the proton interaction region for fully reconstructed final states, and with a slightly wider distribution of impact parameters for partially reconstructed final states (see Fig.~\ref{fig:HS_Performance_FullyReconstructed}).

\subsubsection{Tracking performance}

A key goal of the SHiP experiment is to determine the mass of a potential hidden sector candidate with a high degree of accuracy. The mass resolution of an HNL candidate with mass 1\gevcc reconstructed through the decay \linebreak HNL$\to\mu^\pm\pi^\mp$ is shown in Fig.~\ref{fig:HS_Performance_FullyReconstructed}. The dependence of the mass resolution as a function of the mass of the HS particle for decays to both fully and partially reconstructed modes is also shown.

Accurate vertex and momentum resolution are also critical to suppress backgrounds. As HS  particle candidates originate from the target, the impact parameter of the reconstructed candidate with respect to the proton interaction region (IP$_{\rm target}$) offers excellent discriminating power against backgrounds.  In addition, signal tracks originate from a common vertex in contrast to backgrounds arising from random combinations of tracks. Therefore, the distance of closest approach (DOCA) between tracks can also be used to suppress backgrounds. The distributions of IP$_{\rm target}$ and DOCA for typical signals and combinatorial background are shown in Fig.~\ref{fig:HS_Performance_FullyReconstructed}.




\subsubsection{Background rejection}
\label{sec:bkg_rejection}

In order to maximise SHiP's sensitivity, the background in the HSDS decay volume should be reduced to a negligible level. A common set of loose selection criteria are applied to all HS particle searches to suppress the background (Table~\ref{tab:selection}). The selection criteria preserves close to 100\% of the signal efficiency in fully reconstructed modes, as can been seen with the example of 1\gevcc HNLs in Fig.~\ref{fig:HS_Performance_FullyReconstructed}.

\begin{table*}
\centering
 \begin{tabular}{l r} 
 {Requirement} & {Value}  \\ [0.5ex] 
 \hline\hline
 Track momentum & $>$\,1.0\gevc  \\ 
 Track pair distance of closest approach & $<$\,1\,cm  \\
 Track pair vertex position in decay volume & $>$\,5\,cm from inner wall \\
 Impact parameter w.r.t. target (fully reconstructed) & $<$\,10\,cm  \\
 Impact parameter w.r.t. target (partially reconstructed) & $<$\,250\,cm  \\
 \hline
 \end{tabular}
 \caption{Pre-selection criteria used for the background rejection and the sensitivity estimates in the analysis of HS particle decays.}
\label{tab:selection}
\end{table*}

There are three main sources of background that can mimic the signature of HS particles: random combinations of residual muons within the same proton spill, muon DIS and neutrino DIS. The background from cosmics has been proven to be negligible.

\noindent {\bf $\bullet$ Muon combinatorial:} The expected rate of residual muons that enter the fiducial volume of the HSDS, or back-scatter in the cavern and traverse the SHiP spectrometer, is \linebreak 26.3$\pm$3.6\,kHz. After applying the acceptance and selection cuts listed in Table~\ref{tab:selection}, together with a cut of <25 on the sum of hits in all tracking stations and no hit in the SBT/UBT, about $10^8$ pairs of muons are expected for the partially reconstructed topology in the lifetime of the experiment. Under the assumption of a flat time structure for the 1\,s proton spills, these are suppressed to a level of $10^{-2}$ muon pairs in $2\times 10^{20}$~protons on target by requiring the muons to coincide in a time window of 340\,ps, corresponding to >\,2.5 times the time resolution of the HSDS timing detector. In reality, non-uniformity in the spill structure and the probability of this background can be measured by relaxing the timing criterion. A fast high-resolution spill structure monitor is also under study. The information from the monitor will be recorded with the data to have a continuous measure of this background probability. At the same time, significant progress has also been made in the context of BDF to improve the SPS spill structure, and studies of new techniques are underway. 

\noindent {\bf $\bullet$ Muon DIS:} Muons interacting inelastically in the floor and walls of the cavern, and in the material upstream of the vacuum vessel, can produce particles that enter the decay volume and mimic the signal. 
We expect about $2\times 10^8$ muon DIS interactions in the proximity of the vacuum vessel for $2\times 10^{20}$ protons on target. Samples of background corresponding to the expected number of DIS interactions
have been generated (see Section~\ref{sec:simulation}). No events remain after applying the pre-selection in Table~\ref{tab:selection} and the detector information from the SBT and the UBT. To further investigate the background suppression it is assumed that the background taggers' veto and the pointing criteria factorize. This results in an upper limit of  $6\times 10^{-4}$ expected background events for $2\times 10^{20}$ protons on target.

\noindent {\bf $\bullet$ Neutrino DIS:} The dominant source of this background comes from neutrino DIS in the proximity of the decay volume, roughly corresponding to $3.5\times 10^7$ interactions from $2\times 10^{20}$ protons on target. In order to avoid irreducible background from neutrinos interacting with the air molecules inside the vessel, a level of vacuum below $10^{-2}$ bar is necessary. The background from neutrino scattering in the floor and the walls of the cavern was studied and found to be negligible. 
A large sample, corresponding roughly to seven times the planned proton yield of $2\times 10^{20}$ protons on target was generated (see Section~\ref{sec:simulation}). 
By applying the selection cuts listed in Table~\ref{tab:selection} together with the background tagger information and timing, we expect <\,0.1 background events for the fully reconstructed signal and 6.8 background events for the partially reconstructed signal. This background consists of photon conversions in the material. It can be easily eliminated by requiring an invariant mass of the pair to be larger than 100\,MeV/c$^2$.

%
%
As Table~\ref{Tab:bkgs} shows, in the design of the SHiP experiment, much attention has been given to the identification of possible sources of background
and to the means to reduce the background with a high level of redundancy for a broad spectrum of searches for visible decays
of Hidden Sector particles. The redundancy of the selection criteria also allows determining the background  directly from data and, in case of signal evidence, to perform cross checks that minimise the probability of false positives. 
\begin{table}
\centering
 \begin{tabular}{l r r} 
  Background source & Expected events  \\  
  \hline\hline
  Neutrino DIS & $<0.1$ (fully) / $<0.3$ (partially)\\ 
  Muon DIS (factorisation) & $<\,6\times 10^{-4}$ \\
  Muon combinatorial & $1.2\times 10^{-2}$\\
  \hline
     \end{tabular} 
     \caption{Expected background in the search for HS particle decays at 90\%\,CL for $2\times 10^{20}$ protons on target after applying the pre-selection, the timing, veto, and invariant mass cuts. The neutrino-induced background is given separately for fully and partially reconstructed background modes.}
\label{Tab:bkgs}
\end{table}

\section{Conclusions}
\label{sec:conclusions}


While the energy frontier is investigated at the LHC and the precision frontier is pursued at LHCb, Belle II, NA62 and elsewhere, the intensity frontier remains under-explored. In the wake of the Higgs discovery, the SHiP collaboration identified a unique opportunity to pursue a new path of direct searches for a large class of feebly interacting particles, including light dark matter, by developing a novel type of beam-dump facility at the CERN SPS. The SPS is currently under-exploited and can provide a yield of protons to explore the interesting parameters space of HS particles that greatly surpasses existing facilities. 

SHiP has since instigated a number of pioneering developments that make it possible to construct a large-scale, high-precision detector operating in beam-dump mode with 4$\times 10^{19}$ protons per year at 400\,\gevc and in an environment of extremely low background. The detector setup is capable of reconstructing the decay vertex of a HS particle, measuring its invariant mass and providing particle identification of the decay products. Detailed full MC simulations have demonstrated that SHiP can suppress the expected background level in the search for visible decays to below 0.1 event at 90\% C.L.  in $2\times 10^{20}$ protons on target, equivalent to five years of operation at nominal intensity. It has also sensitivity to explore new parameter ranges of light dark matter and tau neutrino physics through scattering signatures in a dedicated set of sub-detectors with an emulsion target. This puts SHiP in an outstanding position world-wide to make a break-through in the range of particle masses and couplings that is not accessible to the energy and precision frontier experiments, and potentially find the particles that lead to neutrino masses and oscillations, explain the baryon asymmetry of the universe, and shed new light on the properties of dark matter. 

All of SHiP’s sub-systems have addressed the most relevant technological challenges and undertaken genuine programs of prototyping to validate their performance with beam tests of small scale prototypes. The results have been used in the full simulation of the expected physics performance. In the 2020 Update of the European Strategy for Particle Physics, the preparatory evaluation of experiments complementary to the high energy frontier singled out SHiP at the SPS Beam Dump Facility as a major potential player in the future search for feebly interacting particles (CERN-PBC-REPORT-2018-003 and~\cite{2020NaturePhysicsPBC}). With regards to the cost of the baseline design of the facility, the project could not, as of 2020, be recommended for construction considering the overall recommendations of the Strategy. Hence, with the feasibility of the facility and the detector proven, CERN and the SHiP collaboration are focusing efforts on reviewing the design of the facility and possible alternative locations at the SPS (CERN-SPSC-2022-009), with the aim to significantly reduce the implementation cost, and enable start of construction in CERN's Long Shutdown 3, while preserving the original physics reach. The detector concept developed for the SND is also being deployed at the LHC (CERN-LHCC-2021-003 / LHCC-P-016) in order to make measurements with neutrinos, in particular from charm production, in an unexplored range of energy and a range of pseudorapidity that is inaccessible to the other LHC experiments, as well as search for light dark matter.

\section{Acknowledgements}
All work related to this publication was performed between 2014 and 2020.
The SHiP Collaboration wishes to thank the Castaldo company (Naples, Italy) for their contribution to the development studies of the decay vessel. We also wish to thank the groups from INFN/University of Bologna and Laboratori Nazionali dell'INFN di Frascati for developing the scintillating tile-based option for the hidden sector muon system.
The support from DFG, Germany, is acknowledged.
The support from the National Research Foundation of Korea with grant numbers of \linebreak 2018R1A2B2007757, 2018R1D1A3B07050649, 2018\-R1D\-1A\-1B07050701, 2017R1D1A1B03036042, 2017\-R1A6A3\-A01075752, 2016R1A2B4012302, and 2016R1A6A3A1\-1\-930680 is acknowledged.
The support from the Russian Foundation for Basic Research, grant 17-02-00607, and the support from the TAEK of Turkey
are acknowledged.
The work was carried out with financial support from the Ministry of Education and Science of the Russian Federation in the framework of the Competitiveness Enhancement Program of NUST ‘‘MISIS”, implemented by a governmental decree dated 16th of March 2013, No 2.  
We would like to thank Andrea Merli, Nicola Neri and Marco Petruzzo (Milano University) for making available a complete silicon strip telescope, with readout and reconstruction software, that allowed us to perform test beam measurements with our prototype straws.

\bibliographystyle{epj} %
\bibliography{All}%


\onecolumn
\noindent
\textbf{The SHiP Collaboration}

\noindent
C.~Ahdida$^{45}$,
A.~Akmete$^{49}$,
R.~Albanese$^{15,c,e}$,
J.~Alt$^{7}$,
A.~Alexandrov$^{15,33,35,c}$,
A.~Anokhina$^{40}$,
S.~Aoki$^{19}$,
G.~Arduini$^{45}$,
E.~Atkin$^{39}$,
N.~Azorskiy$^{30}$
% ...

\vspace*{0.5cm}

{\footnotesize \it
\noindent
% ...
$^{7} $Physikalisches Institut, Universit\"{a}t Freiburg, Freiburg, Germany\\
% ...
$^{15}$Sezione INFN di Napoli, Napoli, Italy\\
% ...
$^{19}$Kobe University, Kobe, Japan\\
% ...
$^{30}$Joint Institute for Nuclear Research (JINR), Dubna, Russia\\
% ...
$^{33}$P.N.~Lebedev Physical Institute (LPI RAS), Moscow, Russia\\
% ...
$^{35}$National University of Science and Technology ``MISiS``, Moscow, Russia\\
% ...
$^{39}$National Research Nuclear University (MEPhI), Moscow, Russia\\
$^{40}$Skobeltsyn Institute of Nuclear Physics of Moscow State University (SINP MSU), Moscow, Russia\\
% ...
$^{45}$European Organization for Nuclear Research (CERN), Geneva, Switzerland\\
% ...
$^{49}$Middle East Technical University (METU), Ankara, Turkey\\
% ...
$^{c}$Universit\`{a} di Napoli ``Federico II``, Napoli, Italy\\
% ...
$^{e}$Consorzio CREATE, Napoli, Italy\\
% ...
}



\end{document}